\documentclass{IEEEtran}
\usepackage[hyphens]{url}
\usepackage{graphicx}
\usepackage{xcolor}
\usepackage{colortbl} 
\usepackage{ulem} 
\usepackage{caption}
\usepackage{subcaption}
\usepackage[bookmarks=false]{hyperref}
\usepackage{hyperref}
\hypersetup{colorlinks=true,linkcolor=black,citecolor=blue,filecolor=black,urlcolor=blue}
\usepackage{algorithm}
\usepackage{algorithmicx} 
\usepackage{algpseudocode}
\usepackage{stmaryrd}

\algnewcommand\algorithmicforeach{\textbf{for each}}
\algdef{S}[FOR]{ForEach}[1]{\algorithmicforeach\ #1\ \algorithmicdo}
\algblock{Input}{EndInput}
\algnotext{EndInput}
\algblock{Output}{EndOutput}
\algnotext{EndOutput}
\newcommand{\Desc}[2]{\State \makebox[2em][l]{#1}#2}

\definecolor{headcolor}{gray}{0.9}

\begin{document}

\title{Evaluation of pilot jobs for Apache Spark\\ applications on HPC clusters}
\author{
    \IEEEauthorblockN{
        Val\'erie Hayot-Sasson and Tristan Glatard \\
        Department of Computer Science and Software Engineering, Concordia University \\
        Montr\'eal, Qu\'ebec, Canada
    }
    \IEEEauthorblockA{}
}
\maketitle

\begin{abstract}
    Big Data has become prominent throughout many scientific fields and, as
    a result, scientific communities have sought out Big Data frameworks to
    accelerate the processing of their increasingly data-intensive
    pipelines. However, while scientific communities typically rely on
    High-Performance Computing (HPC) clusters for the parallelization of
    their pipelines, many popular Big Data frameworks such as Hadoop and
    Apache Spark were primarily designed to be executed on dedicated
    commodity infrastructures. This paper evaluates the benefits of pilot
    jobs over traditional batch submission for Apache Spark on HPC
    clusters. Surprisingly, our results show that the speed-up provided by
    pilot jobs over batch scheduling is moderate to inexistent (0.98 on
    average) despite the presence of long queuing times. In addition, pilot
    jobs provide an extra layer of scheduling that complexifies debugging
    and deployment. We conclude that traditional batch scheduling should
    remain the default strategy to deploy Apache Spark applications on
    HPC clusters.

\end{abstract}

\section{Introduction}

Pilot jobs, also known as dynamic resource provisioning or glide-in
scheduling, are a widely-used technique to address infrastructure
heterogeneity, variable task queuing times, fine task granularity, and node
failures on distributed computing infrastructures. Made popular by software
projects such as Condor~\cite{thain2005distributed} and DIRAC~\cite{casajus2010dirac}, they
became critical to grid computing, greatly improved the performance of HPC
clusters, and enabled multi-cloud executions. 

In contrast to static resource provisioning, pilot jobs are submitted to
the infrastructure separately from the application, to provision resources
on which application tasks will eventually be scheduled. Large pools of
resources can thus be created, shielding applications from the underlying
queuing times, failures, and other idiosyncracies. A variety of frameworks
now rely on pilot jobs, including RADICAL-Pilot~\cite{merzky2015radical},
and recent versions of the Pipeline System for Octave and Matlab
(PSOM)~\cite{bellec2012pipeline}. The survey in~\cite{turilli2018comprehensive} reviews
the current pilot-job systems.

In this paper, we study the use of pilot jobs to run scientific
applications with Apache Spark~\cite{zaharia2016apache} on shared HPC
clusters. The method currently recommended for this purpose involves batch
requesting all the
necessary resources and launching a Standalone Spark cluster once the
resources have been allocated. This is, for instance, the method used on
Compute Canada,
 our national computing infrastructure. With pilot
jobs, rather than requesting all the resources at once, a Spark cluster is
launched with a subset of the resources, and is expanded as more resources
get allocated. Consistently with previous examples of pilot job
deployments, we hypothesize that this strategy would reduce queuing times
by (1) fragmenting resource requirements, and (2) allowing shorter walltime
estimates. While there have been some efforts on implementing pilot jobs
for Apache Spark~\cite{luckow2016hadoop}, research is limited and
none of the them detail their quantitative effect.

Apache Spark is a popular Big Data framework, commonly used in both
industrial and academic settings. Although it is a Scala-based framework,
it also has APIs for Java, Python (PySpark) and R. Spark's Resilient
Distributed Dataset (RDD) abstraction enabled in-memory processing of
pipelines by co-locating tasks and data, which provided important
performance improvements compared to its predecessor Hadoop
MapReduce~\cite{dean2008mapreduce}. Through the use of RDDs, it also became possible
to execute iterative workflows -- something not easily doable in older
frameworks. Schedulers for Spark include its built-in standalone schedule,
Yet Another Resource Negotiator (YARN~\cite{apache13yet}), and Mesos~\cite{hindman2011mesos}. As a
result of the sustained growth in data volumes, Apache Spark and other
Big Data engines are increasingly used for scientific applications,
including in neuroimaging, our primary field of
interest~\cite{boubela2016big,mehta2017comparative,freeman2014mapping}.

Big Data frameworks were designed with dedicated commodity infrastructure
in mind, and, with the exception of Dask~\cite{rocklin2015dask}, do not support batch
HPC schedulers such as the \href{http://www.univa.com/products/}{Oracle Grid Engine (OGE)},
\href{https://slurm.schedmd.com/}{Slurm} and \href{https://www.adaptivecomputing.com/products/torque/}{TORQUE}, which
are commonly available to scientists. Therefore, to run Big Data
applications on HPC schedulers, it is also necessary to start an overlay
cluster that will schedule application tasks on resources provisioned
through batch schedulers. Our experiments quantify the effect of pilot jobs
when combined with such an overlay cluster.

To summarize, our paper
makes the following contributions:
\begin{itemize}
\item We present SPA, a lightweight pilot-job framework to run Apache Spark
applications on HPC clusters.
\item We compare the makespan of a typical neuroimaging Big Data
application with and without pilot jobs on two different HPC clusters and
in different conditions.
\item We describe a simple performance model to validate that the observed
differences come from variations in queuing times rather than other
factors.
\end{itemize}
Our methods, including infrastructure, job templates, application and
performance model are described in Section~\ref{sec:methods}.
Section~\ref{sec:results} presents our results which are discussed in
Section~\ref{sec:discussion}. Section~\ref{sec:conclusion} concludes on the 
relevant of pilot jobs for Apache Spark applications on HPC clusters.

\section{Materials and Methods}\label{sec:methods}

    The application, templates, configuration files, benchmarks and
    analysis scripts are publicly available and can be found in our Spark
    Pilot-job scheduler for HPC Applications (SPA) repository at:
    \href{https://github.com/big-data-lab-team/spa}{https://github.com/big-data-lab-team/spa}.
    Links to the processing engines and processed data are provided in the
    text.
    
    \subsection{Infrastructure}
    All experiments were conducted on the Cedar and B\'eluga HPC computing
    clusters made available by \href{https://www.computecanada.ca}{Compute
    Canada} through \href{https://www.westgrid.ca}{WestGrid} and
    \href{http://www.calculquebec.ca}{Calcul Qu\'ebec}. Both clusters are
    accessible through the Slurm batch scheduler and Lustre parallel file
    system~\cite{schwan2003lustre}. The Cedar cluster has a total of 1542
    nodes with a total of 58,416 CPU cores. Available memory on a Cedar
    node can range from 125 to 3022~GB. Standard nodes are equipped with
    either 2x Intel E5-2683 v4 Broadwell @ 2.1~Ghz (32 cores total) or 2x
    Intel Platinum 8160F Skylake @ 2.1Ghz (48 cores total) CPUs and 2 x
    480~GB SSD. All nodes and temporary storage on Cedar are connected by an
    Intel OmniPath (version 1) with 100Gbit/s bandwidth.

    B\'eluga, on the other hand, is a smaller cluster with 872 available
    nodes. Node memory can range between 92 to 752~GB, with the most common
    node type having 186G. All nodes contain 2 x Intel Gold 6148 Skylake @
    2.4~Ghz (40 cores/node) CPU and are connected to each other with a
    100~Gb/s Mellanox Infiniband EDR network. Each non-GPU
    node type contains one 480~GB SSD. 

    It is important to note that these clusters were used in production,
    that is, concurrently with other users. This allowed us to test the
    added-value of pilot scheduling in different realistic conditions of queuing
    times. At the time of our experiments, B\'eluga had recently been
    commissioned, resulting in low usage and shorter queuing times overall.
    Conversely, Cedar had been operating for a few years, resulting in
    higher usage and longer queuing times. We repeated our experiments multiple 
    times in each configuration to capture queuing time variability.

    \subsection{Spark Configuration}

    There are three possible cluster managers available to use in Spark applications:
    the Standalone cluster manager, YARN and Mesos. The Standalone cluster manager is the
    most basic cluster manager available and is packaged directly with Spark. YARN is a
    resource manager designed for the resource management of Hadoop applications. However,
    it can be used with different types of applications as well. In contrast, Mesos was 
    designed to manage the resources for a variety of applications. As such, it incorporates
    strategies that are more efficient for use with different frameworks making it possible
    to be used as an HPC scheduler. Due to Standalone's lightweight nature and our intent
    to start Spark clusters per application, we focus on Standalone mode.

    A Spark cluster is made up of three components: the cluster manager, workers and
    driver. The cluster manager, also known as the Spark master in Standalone mode, is responsible for the
    resource provisioning within the cluster.  As it is the resource manager for a given cluster,
    the cluster manager is not application specific. The driver requests resources from the master
    to launch its tasks. The master then dispatches the requests to workers with
    the necessary resources available to start
    executors. The driver may subsequently communicate with the executors to schedule and launch its tasks. Both the driver and
    the executors are application specific and connected directly to the master. 

    Spark has many configuration options available for dynamic resource provisioning.
    Its standalone cluster provides two deployment modes: client and cluster. The client
    deploy mode executes the driver within the \texttt{spark-submit} process as
    a client to the cluster. The cluster deploy-mode, however, runs the driver within
    a worker process. Such a configuration is practical when the driver program
    is deployed on a machine not co-located in the cluster to reduce network 
    latency between workers and driver. The cluster deploy mode also has the \texttt{supervise}
    feature which allows the driver to be restarted in case of failure.
    This is particularly useful for
    pilot-based overlay clusters where the running driver may fail due to walltime expiration.
    Using the Standalone cluster manager, client mode is supported in all APIs, whereas cluster mode is
    only available for applications using the Scala, Java and R APIs.

    With HPC clusters, client mode is not necessarily possible on the login node as the
    driver client may require more resources than permitted. Moreover, due to cluster
    security policies, it may be difficult or impossible, to execute the driver from
    an external computer, such as a personal laptop. Therefore, for both batch and pilot
    scheduling, it is necessary to launch the driver from within the Slurm jobs. That is, in cluster mode.

    Spark also provides master fault tolerance through single-node recovery, wherever a shared
    filesystem is available. To enable this, a recovery folder must be set. Should the master
    fail, a new master will be able to takeover based on the information available in the recovery
    folder.

    In long running applications, it is necessary to be able to recover application execution from
    last successful state. Spark provides a checkpointing mechanism in order to ensure that the application
    is fault-tolerant to any kind of non application-related failure, such as a system failure.

    \subsection{SPA system design}

    There is no shortage of scripts available online for starting a Spark cluster on HPC, 
    see for instance \href{https://github.com/NIH-HPC/spark-slurm}{spark-slurm},
    \href{https://www.sherlock.stanford.edu/docs/software/using/spark}{scripts} provided
     by Stanford's Sherlock cluster, \href{https://sparkhpc.readthedocs.io}{sparkhpc},
     or \href{https://www.osc.edu/~troy/pbstools/man/pbs-spark-submit}{pbs-spark-submit}.
    However, all these scripts start up a cluster within a single batch call. At most, there are two calls, one for
    cluster startup,
    and the other, to launch the application. None of these use any kind of pilot-scheduling approach, nor do they discuss the
    effects of queuing time and how they might want to adjust it in the
    case where the driver is started separately from the batch job to start
    the cluster. Even in the case of RADICAL-Pilot, as per some available
    tutorials for using the pilot-job application with
    Spark (see
    \href{https://github.com/radical-cybertools/pilot-streaming/blob/master/examples/Pilot-Streaming-GettingStarted.ipynb}{here}
    and \href{https://github.com/radical-cybertools/MIDAS-tutorial/blob/master/pilot/Pilot-Spark.ipynb}{here}),
    it appears that the Spark cluster is started within a single pilot and
    the application is submitted, in a separate process, to that pilot. Due
    to the fact that existing solutions mainly rely on a single batch call
    to startup the Spark cluster and do not appear to be able to submit
    multiple smaller jobs to improve queuing times, we have decided to
    implement our own solution for the sake of our experiments.

    Two different job templates were developed to implement the two main
    conditions compared in our experiments: the batch submission
    template and the pilot submission template. The batch submission template was
    inspired by the template provided by Compute Canada to launch Spark applications 
    on Slurm, \href{https://docs.computecanada.ca/wiki/Apache\_Spark/en}{available here}.
    The template operates as follows: certain resource requirements are requested by
    the user (e.g. walltime, amount of memory per node, number of CPUs per task, number 
    of nodes and number of tasks per node). Once these resources are allocated, a 
    master is started on one of the requested node resources. Then, after the master
    has successfully started, the workers are started on all nodes. Multiple worker
    instances are started on a single node by setting the \texttt{SPARK\_WORKER\_INSTANCES}
    environment variable to the number of tasks per node. Each worker is given as many
     cores as specified by the user in the Slurm resource allocation request.
    After both the masters and the workers have successfully started, the driver is finally
    started. The amount of memory given to each executor corresponds to 95\% of the 
    available memory on the node, to allow for offheap space. 
    The Spark deploy mode selected for the batch template is client mode. We selected this
    deploy mode over cluster mode as driver recovery would not be required in batch in the
    case of walltime expiration, as all the allocated resources will expire at the same time.
    Moreover, many scientific applications are written in Python, which cannot use cluster mode using
    the Standalone resource manager. Should pilot scheduling using cluster mode significantly reduce
    queuing time compared to batch scheduling using client mode, it would provide enough justification to extend
    Spark's Standalone scheduler to include cluster mode for PySpark applications.

    The pilot submission template is similar to that of the batch template, although,
    each pilot will start its own Spark master and worker. However, there will only be one
    pilot which will start the driver process. The pilot selected to start the driver is the first one
    that attempts to do so by way of lockfile.
    The reason for 
    which each pilot starts its own master is to ensure the fault tolerance of the 
    masters. In this configuration, should the active master be killed, one
    of the stand-by masters can takeover and the application may be able to 
    resume if single-node recovery is set. Such a configuration is particularly favourable in pilot scheduling 
    scenarios as node failures may be more frequent due to walltime expiration.
    Additionally, the Spark deploy mode of the driver was selected to be cluster deploy mode.
    This would not only allow the driver to be executed directly on one of the workers,
    but also allow us to make the driver fault-tolerant through the
    \texttt{supervise} mode, which is only available in cluster deploy. As
    with the masters, it is particularly important to have a fault-tolerant
    driver in pilot-scheduling scenarios due to possible walltime
    expiration.
    Should pilots be idle for a certain duration, the
    pilots will terminate themselves such as to not hog resources.

    Although we start a master on each pilot, we have not incorporated the master recovery function
    for our experiments. This feature is, however, important and should be incorporated in pilot schedulers
    were any nodes can fail by design.

    Due to the differences in deploy modes between batch and pilot submission, 
    batch will always inevitably have one more worker than pilot. This is because
    in cluster deploy, which pilot uses, the driver occupies a worker, whereas in
    client deploy, the driver is separate from any worker.

    Both of these Slurm templates are launched within a Python application called
    \textit{SPA}. The templates are used in conjunction with JSON configuration 
    file and passed to the \href{https://github.com/brentp/slurmpy}{SlurmPy} library within \textit{SPA}.
    The \textit{SPA} application, all the while ensures that all is preconfigured
    correctly before passing it to SlurmPy. It also ensures that enough pilots are
    launched, maintains track of the running/queued pilots, and launches additional
    pilots if there are less pilots than requested by the user in the Slurm queue.

    \subsection{Application}
        \begin{algorithm}\caption{Incrementation}\label{alg:incrementation}              
            \begin{algorithmic}[1]                                                       
                \Input                                                                       
                    \Desc{$x$}{a sleep delay in seconds}                                         
                    \Desc{$n$}{a number of iterations}                                           
                    \Desc{$C$}{a set of image chunks}                                            
                \EndInput                                                                    
                \ForEach{$chunk$ in $C$}                                                      
                    \State read $chunk$                                        
                    \For{$i \in \llbracket 1, n \rrbracket$}                                                         
                        \State $chunk\gets chunk+1$                                              
                        \State sleep $x$                               
                    \EndFor                                                                      
                    \State write $chunk$                                            
                \EndFor                                                                      
            \end{algorithmic}                                                                
        \end{algorithm}
    To determine the added value of pilot scheduling over batch scheduling of Spark
    applications, we required a Spark application operating on a large dataset with
    an important processing time to emulate what would be the average requirements
    of a scientific Spark application. For this, we created a synthetic application 
    that would process the Big Brain~\cite{amunts2013bigbrain}, a 76~GB 3D histological
    image of a human brain. The algorithm is a chain of map transformations that
    at each transformation increment the voxels of the image by 1 (see Algorithm~\ref{alg:incrementation}).
    We chose such a synthetic algorithm
    as the focus of our experiments is pilot scheduling and not the application in
    itself. Furthermore, this algorithm enabled us to have control over the task duration
    which was representative of scientific applications. Additionally, it was important
    that the overall application duration did not vary between the different levels of
    parallelism within our experiments. Being able to adjust the task duration based on
    level of parallelism allowed us to achieve this.

    Spark cluster fault-tolerance is important in determining the suitability of
    pilot-jobs for Spark applications on HPC. Executing jobs on a shared cluster may
    result in a variety of failures. Pilot jobs may be more likely to fail as a result of
    underestimation of resources (e.g. walltimes) and therefore should be fault-tolerant
    to master, worker and driver failures. Although built-in fault-tolerance was not
    investigated in this paper, we want to ensure that the environment is set up in such
    a way that Spark's fault-tolerance configuration could easily be set should our 
    experiments return favourable results for pilot-jobs.
    Fault-tolerance of the driver is only possible in cluster deploy mode, however,
    when using Spark's Standalone scheduler, this mode is not available for Python 
    applications. It is for this reason that our synthetic application is written
    in Scala. Nevertheless, cluster deploy mode is possible for Python application
    using YARN or Mesos schedulers.

    \subsection{Performance model}

    The makespan of an application is defined as the total duration
    between the submission time of the first application task, and the
    completion time of the last application task. It includes any
    scheduling time, queueing time, data transfer time, and any other
    overhead.
    
    Assuming a divisible load, i.e., the application can be divided in any
    number of tasks, the makespan can be written using the following
    expression, which holds for both batch and pilot execution modes:
    \begin{equation}
        M = \frac{C}{W} \label{eq:mcw}
    \end{equation}
    where:
    \begin{itemize}
        \item $M$ is the makespan of the application
        \item $C$ is the total CPU time of the application
        \item $W$ is the \emph{average} number of Spark workers throughout the execution
    \end{itemize}
    The average number of workers $W$ allows the model to take into account
     variable queuing times. It is computed as follows:
    \begin{equation}
        W = \frac{1}{M}\int_0^M{w(t)dt}\label{eq:avgw}
    \end{equation}
    where $w(t)$ is the number of workers available at time $t$. When the
    application is not subject to any scheduling or queuing time, the
    average number of workers equals the number of workers requested. 

    Therefore, assuming a fixed total CPU time, the relation
    between batch and pilot jobs can be represented as:
    \begin{equation}
        \frac{M_{batch}}{M_{pilot}} = \frac{W_{pilot}}{W_{batch}}\label{eq:makespancomp}
    \end{equation}
    where:
    \begin{itemize}
        \item $M_{batch}$ is the makespan of the batch application
        \item $M_{pilot}$ is the makespan of the pilot application
        \item $W_{pilot}$ is the average number of workers of the pilot application
        \item $W_{batch}$ is the average number of workers of the batch application
    \end{itemize}
    We will use this relation to discuss our results later on. It
    corresponds to an ideal case where no data or other overhead is
    present: only queuing times are included, through the integration in
    Equation~\ref{eq:avgw}.

    \subsection{Added value of pilot scheduling}
        \begin{table}                                                                    
            \centering                                                                       
            \begin{tabular}{c|c|c|c}                                                             
            \rowcolor{headcolor}                                                             
            Configuration & RAM (GB) & Tasks & Cores per task\\                               
            \hline                                                                           
            1 & 112 & 16 & 1\\                                               
            2 & 224 & 32 & 1\\                                               
            3 & 336 & 48 & 1\\
            4 & 448 & 64 & 1\\
            \end{tabular}                                                                    
            \setlength{\belowcaptionskip}{-10pt}                                             
            \caption{Resource configurations}                                                    
            \label{table:dedicatednodes}                                                            
        \end{table} 
           
        \begin{table*}                                                                   
        \centering                                                                       
        \begin{tabular}{c|cccccc}                                                   
          \rowcolor{headcolor}                                                           
          \multicolumn{7}{c}{Configuration 1}\\                      
          \hline                                                                         
          \rowcolor{headcolor}                                                           
          Execution mode & Nodes/job & RAM (GB) & CPUs per task & Tasks/node & Walltime & Task delay (s) \\                             
          \hline
          Batch & 1 & 112 & 1 & 16 & 2h30 & 45 \\
          8 pilots & 1 & 14 & 1 & 2 & 2h30 & 45 \\
          16 pilots & 1 & 7 & 1 & 1 & 2h30 & 45 \\

          \hline                                                                           
          \multicolumn{7}{c}{}\\                                                        

          \rowcolor{headcolor}                                                           
          \multicolumn{7}{c}{Configuration 2}\\                      
          \hline                                                                         
          \rowcolor{headcolor}                                                           
          Execution mode & Nodes/job & RAM (GB) & CPUs per task & Tasks/node & Walltime & Task delay (s) \\                             
          \hline
          Batch & 2 & 112 & 1 & 16 & 2h30 & 90 \\
          8 pilots & 1 & 28 & 1 & 4 & 2h30 & 90 \\
          16 pilots & 1 & 14 & 1 & 2 & 2h30 & 90 \\

          \hline                                                                           
          \multicolumn{7}{c}{}\\                                                        

          \rowcolor{headcolor}                                                           
          \multicolumn{7}{c}{Configuration 3}\\                      
          \hline                                                                         
          \rowcolor{headcolor}                                                           
          Execution mode & Nodes/job & RAM (GB) & CPUs per task & Tasks/node & Walltime & Task delay (s) \\                             
          \hline
          Batch & 3 & 112 & 1 & 16 & 2h30 & 120 \\
          8 pilots & 1 & 42 & 1 & 6 & 2h30 & 120 \\
          16 pilots & 1 & 21 & 1 & 3 & 2h30 & 120 \\

          \hline                                                                           
          \multicolumn{7}{c}{}\\                                                        

          \rowcolor{headcolor}                                                           
          \multicolumn{7}{c}{Configuration 4}\\                      
          \hline                                                                         
          \rowcolor{headcolor}                                                           
          Execution mode & Nodes/job & RAM (GB) & CPUs per task & Tasks/node & Walltime & Task delay (s) \\                             
          \hline
          Batch & 4 & 112 & 1 & 16 & 2h30 & 180 \\
          8 pilots & 1 & 56 & 1 & 8 & 2h30 & 180 \\
          16 pilots & 1 & 28 & 1 & 4 & 2h30 & 180 \\

          \hline                                                                           
          \multicolumn{7}{c}{}\\                                                        
        \end{tabular}                                                                    
        \setlength{\belowcaptionskip}{-10pt}                                             
        \caption{Experimental conditions}
        \label{table:conditions}                                                        
        \end{table*}                                                                                   
        To determine if there are any performance benefits to using pilot over 
        batch scheduling, we needed to compare both strategies given various resource
        requirements. It is expected that large batch requests will stay in the 
        resource queue longer than multiple pilot requests that ultimately use up
        the same amount of resources as each individual pilot requests less resources at
        a time. We therefore used four resource configurations to investigate this
        hypothesis (Table~\ref{table:dedicatednodes}).

       For batch, we requested 1 to 4 dedicated nodes, depending on resource 
       configuration, each with 112~GB of RAM, 1 core per task and 16 tasks per 
       node. On the other hand, for all configurations, we ran our experiments
       with 8 and 16 pilots. Experimental conditions can be seen in Table~\ref{table:conditions}.

       As the focus of our experiments here is to measure the impact pilot scheduling
       has on queueing times, we wanted to ensure that our walltime estimates would
       be consistent for each experimental condition. Therefore, we adjusted task 
       delay based on the maximum level of parallelism to ensure that walltimes would
       not need to be readjusted for each configuration. Given utmost parallelism,
       the sleep delay added would amount to a makespan of 1 hour regardless
       of configuration. The actual processing time of the application, however, 
       would differ between configurations as we did not account for task duration (incrementation time, line
       8 in Algorithm~\ref{alg:incrementation})
       in the sleep delay. Due to slight variability in task durations between experiments, in 
       addition to variability in application parallelism due to queuing times of each pilot Slurm job, 
       we have set the walltime to be 2h30 to ensure experiment completion.

       For each configuration, all three execution modes (batch, 8 pilots,
       16 pilots) were executed in parallel to ensure that the status of
       the HPC cluster was the same when the different execution modes
       were launched. Furthermore, the submission order of the execution
       modes for each configuration was randomized to ensure that Slurm job
       request order could not have affected results. 

       There were 10 repetitions in total for each configuration, and the order
       in which the configurations were launched was also randomized. This was to 
       account for any system variability that can occur, particularly in production
       HPC clusters. 

       Different clusters may have different resource configurations, number of 
       users and scheduling policies. Therefore, we executed all 10 repetitions on
       both Cedar and B\'eluga to determine how much our results differ between
       two distinct clusters.




\section{Results}\label{sec:results}

\begin{figure*}
        \centering
        \begin{subfigure}[b]{0.475\textwidth}
            \centering
            \includegraphics[width=\textwidth]{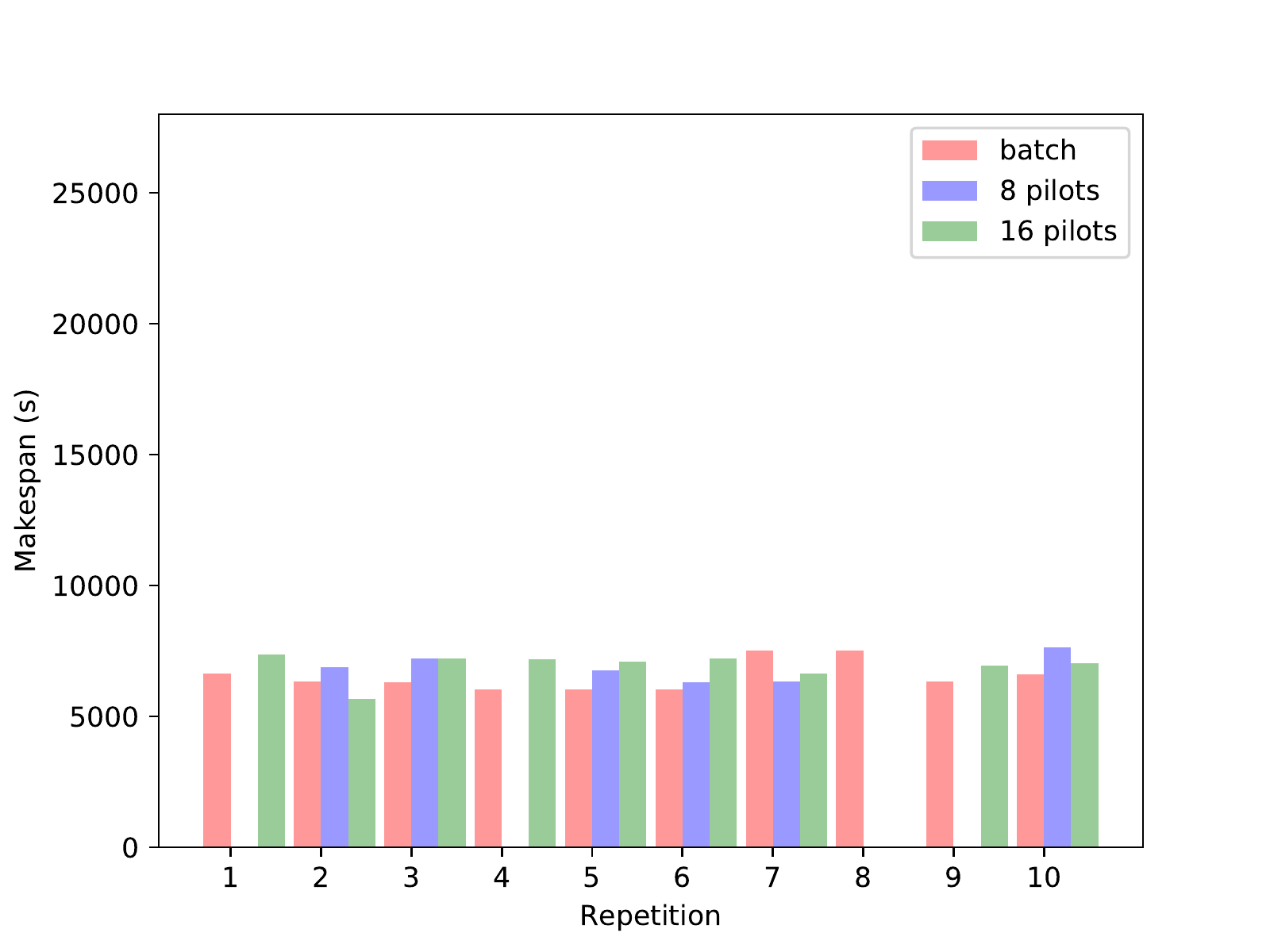}
            \caption[]%
            {{\small Configuration 1}}
            \label{fig:beluga1}
        \end{subfigure}
        \hfill
        \begin{subfigure}[b]{0.475\textwidth}
            \centering
            \includegraphics[width=\textwidth]{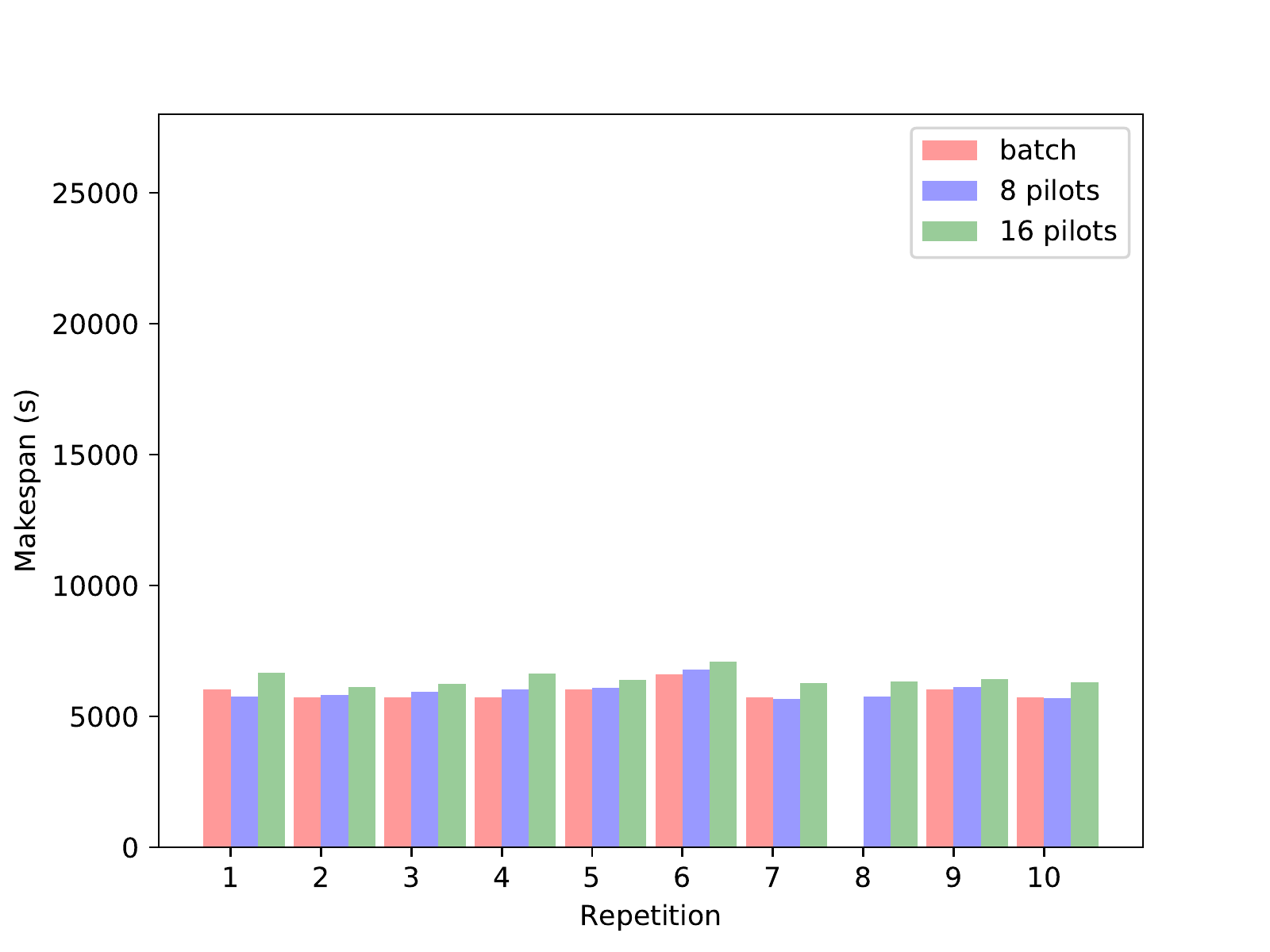}
            \caption[]%
            {{\small Configuration 2}}
            \label{fig:beluga2}
        \end{subfigure}
        \vskip\baselineskip
        \begin{subfigure}[b]{0.475\textwidth}
            \centering
            \includegraphics[width=\textwidth]{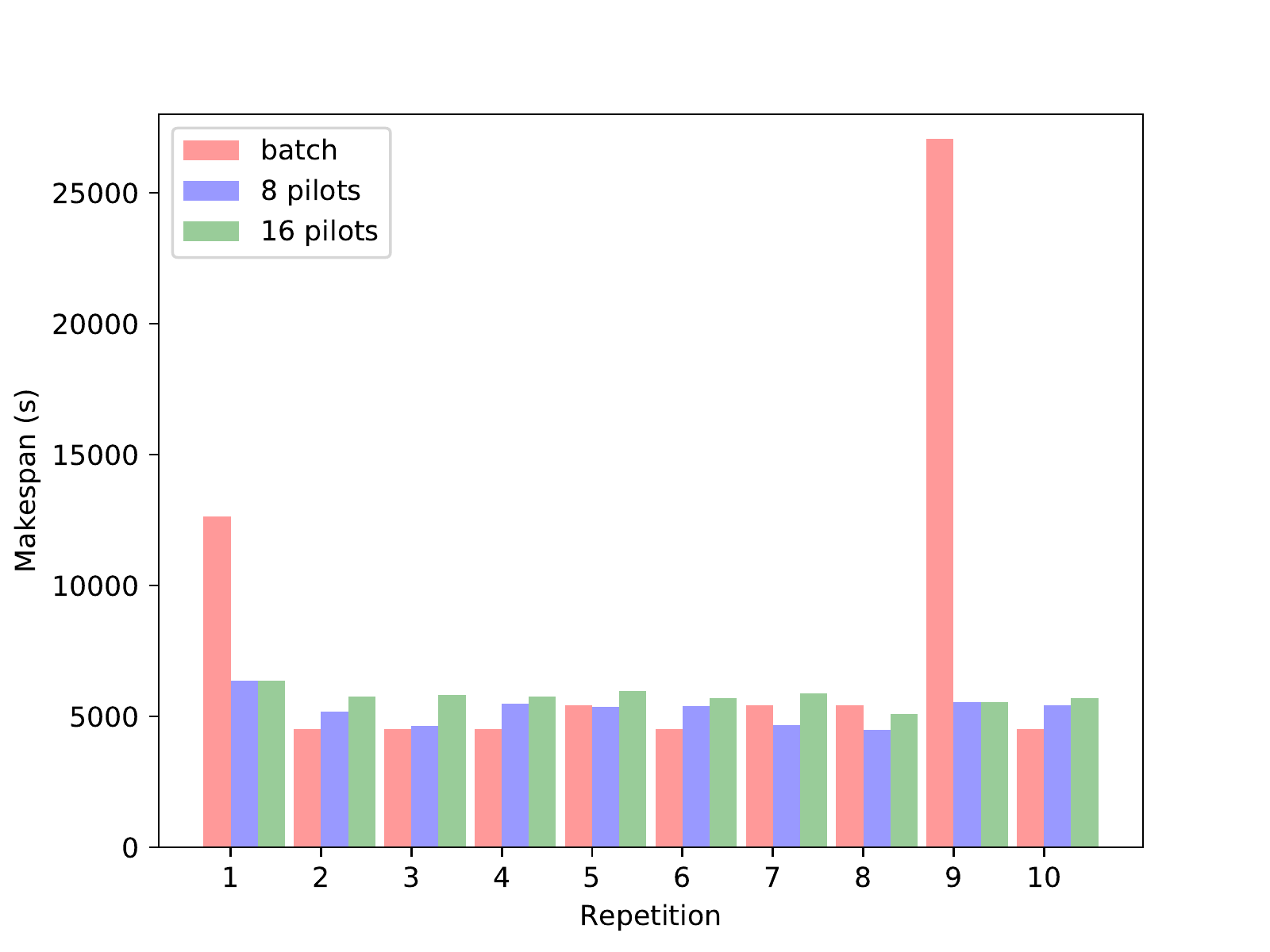}
            \caption[]%
            {{\small Configuration 3}}
            \label{fig:beluga3}
        \end{subfigure}
        \quad
        \begin{subfigure}[b]{0.475\textwidth}
            \centering
            \includegraphics[width=\textwidth]{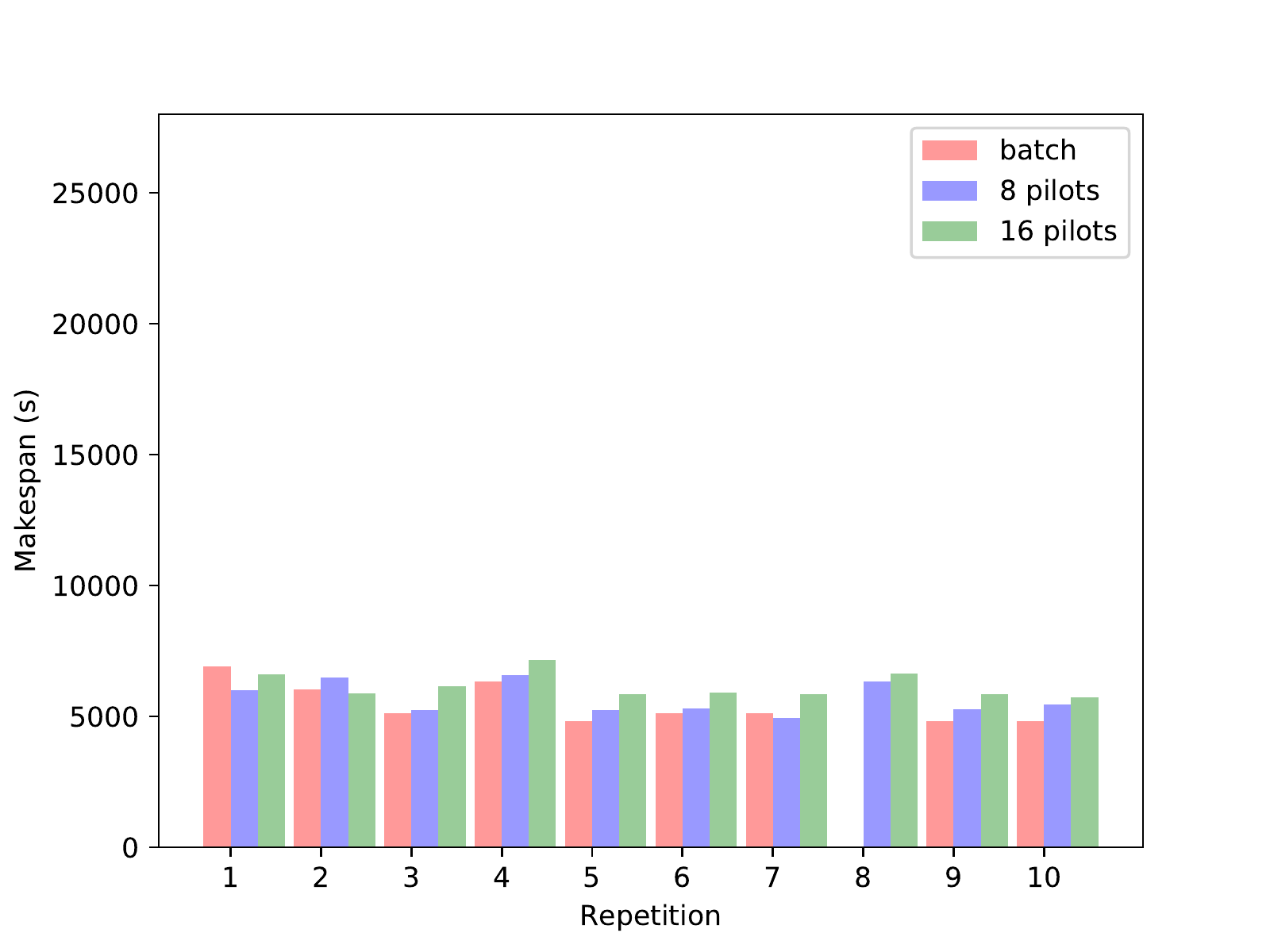}
            \caption[]%
            {{\small Configuration 4}}
            \label{fig:beluga4}
        \end{subfigure}
        \caption[]
        {\small The application makespan on the Beluga cluster for all repetitions.}
        \label{fig:makespansbeluga}
    \end{figure*}

\begin{figure*}
        \centering
        \begin{subfigure}[b]{0.475\textwidth}
            \centering
            \includegraphics[width=\textwidth]{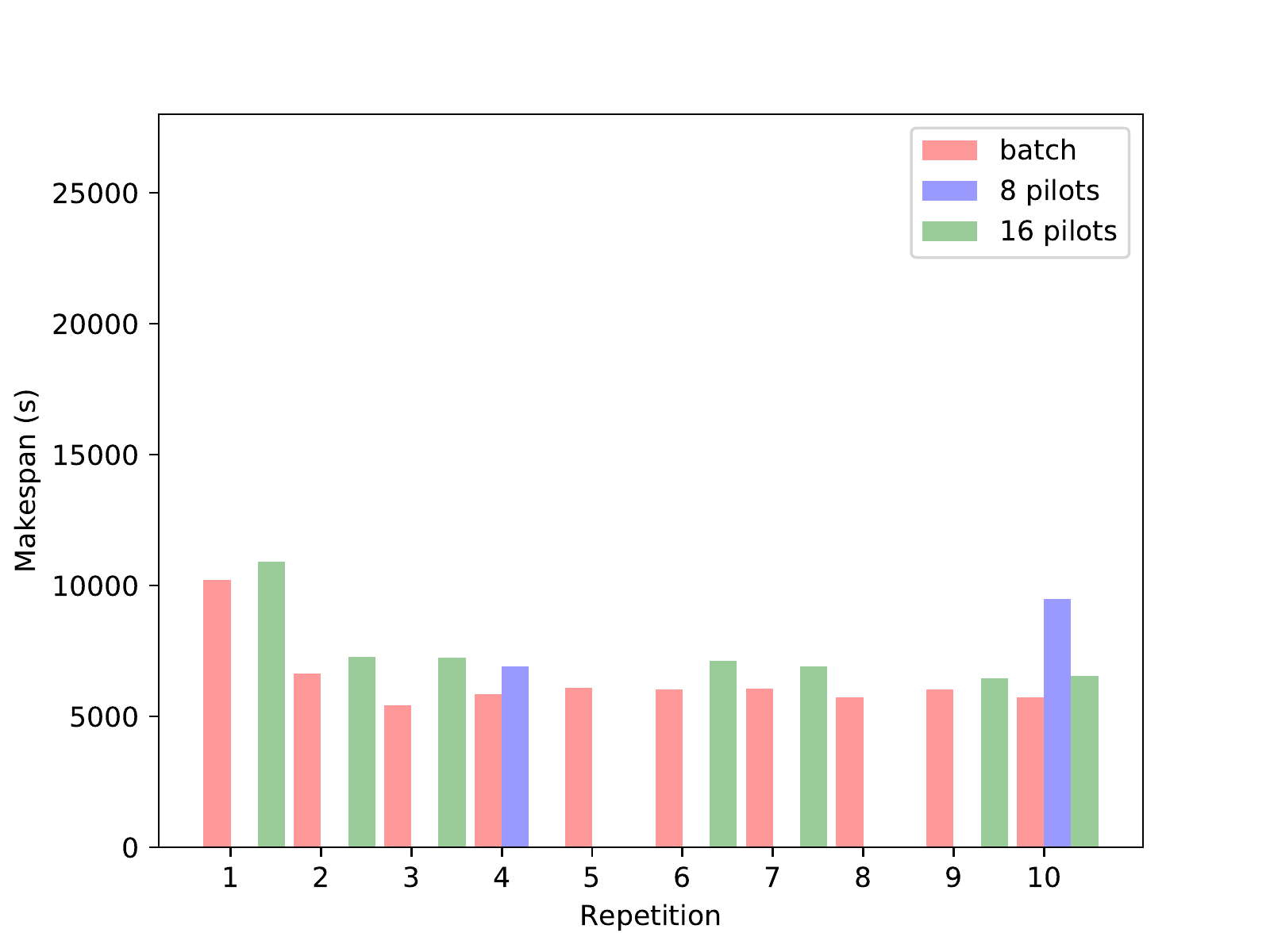}
            \caption[]%
            {{\small Configuration 1}}
            \label{fig:cedar1}
        \end{subfigure}
        \hfill
        \begin{subfigure}[b]{0.475\textwidth}
            \centering
            \includegraphics[width=\textwidth]{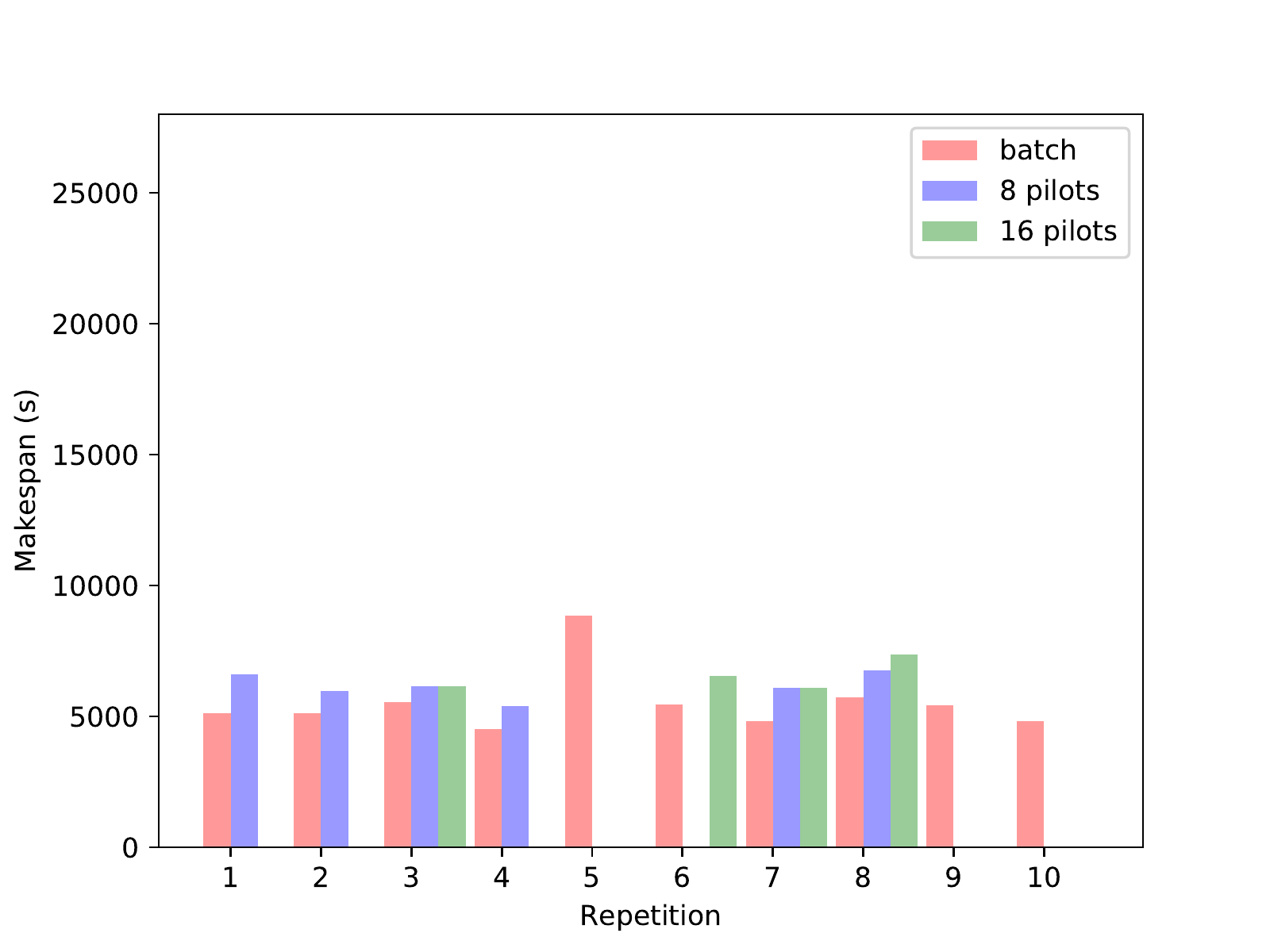}
            \caption[]%
            {{\small Configuration 2}}
            \label{fig:cedar2}
        \end{subfigure}
        \vskip\baselineskip
        \begin{subfigure}[b]{0.475\textwidth}
            \centering
            \includegraphics[width=\textwidth]{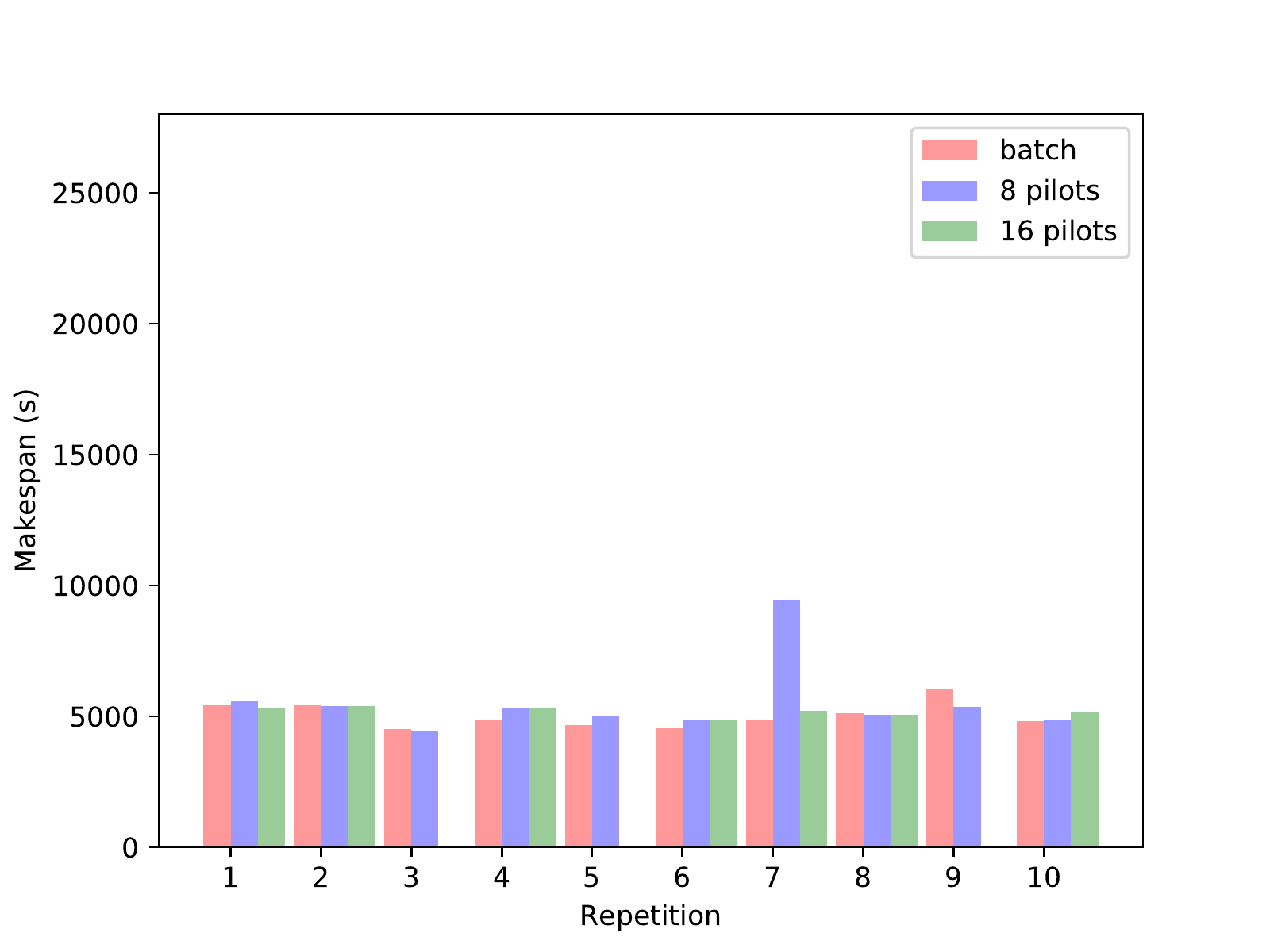}
            \caption[]%
            {{\small Configuration 3}}
            \label{fig:cedar3}
        \end{subfigure}
        \quad
        \begin{subfigure}[b]{0.475\textwidth}
            \centering
            \includegraphics[width=\textwidth]{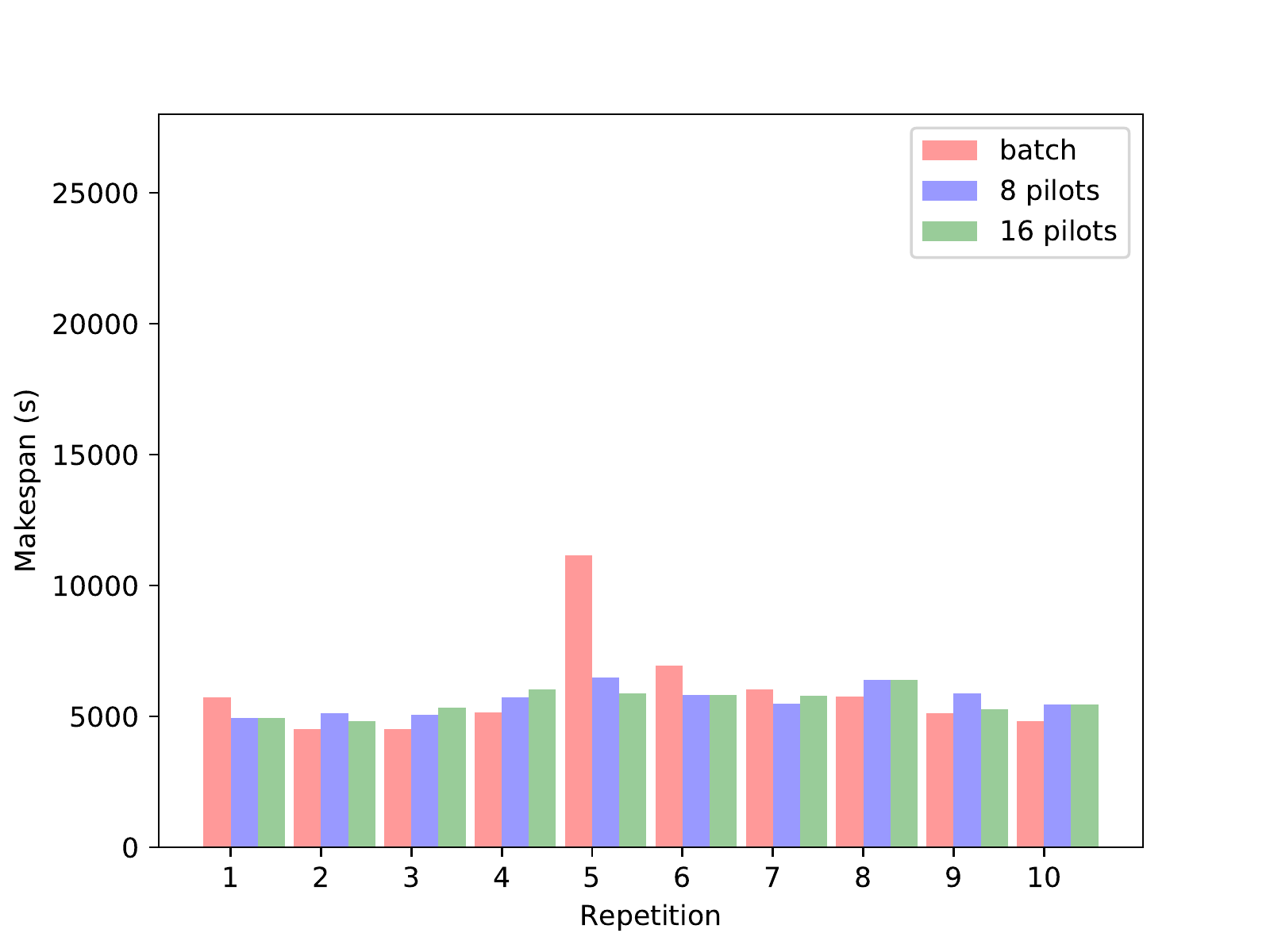}
            \caption[]%
            {{\small Configuration 4}}
            \label{fig:cedar4}
        \end{subfigure}
        \caption[]
        {\small The application makespan on the Cedar cluster for all repetitions.}
        \label{fig:makespanscedar}
    \end{figure*}

As can be seen in both Figures~\ref{fig:makespansbeluga} and~\ref{fig:makespanscedar}, pilots 
generally did not bring any performance improvement to the execution of Spark workloads
on either system. In B\'eluga and Cedar, 16 pilots were generally slower than batch except for a 
few occurrencs. When 16 pilots were faster, it was generally either very slight, or, as expected,
when batch queuing times were very large. The maximum speedup achieved
was approximately 5.6~x for 8 and 16 pilots on B\'eluga and 2.6~x on Cedar. 
It is only in Configuration~3 (Figure~\ref{fig:beluga4}), on B\'eluga,
that we spot any large speedups, however, it generally remains that
pilots are slightly slower, even in Configuration~3. The same can be said about Cedar, however,
it occured with Configuration~4 instead of 3. On average, pilots were up to 0.11~x slower on B\'eluga and
0.17~x slower on Cedar (see Table~\ref{table:speedup}), both of which occurred when using 16 pilots.
The only times when pilots were found to be, on average, faster, was in Configuration~3 on B\'eluga
and Configuration~4 on Cedar where the speedup was found to reach a max of 1.46 for 8 pilots on 
B\'eluga. The missing bars in Figures~\ref{fig:makespansbeluga} and~\ref{fig:makespanscedar} correspond to 
failed executions.

\begin{figure*}
        \centering
        \begin{subfigure}[b]{0.475\textwidth}
            \centering
            \includegraphics[width=\textwidth]{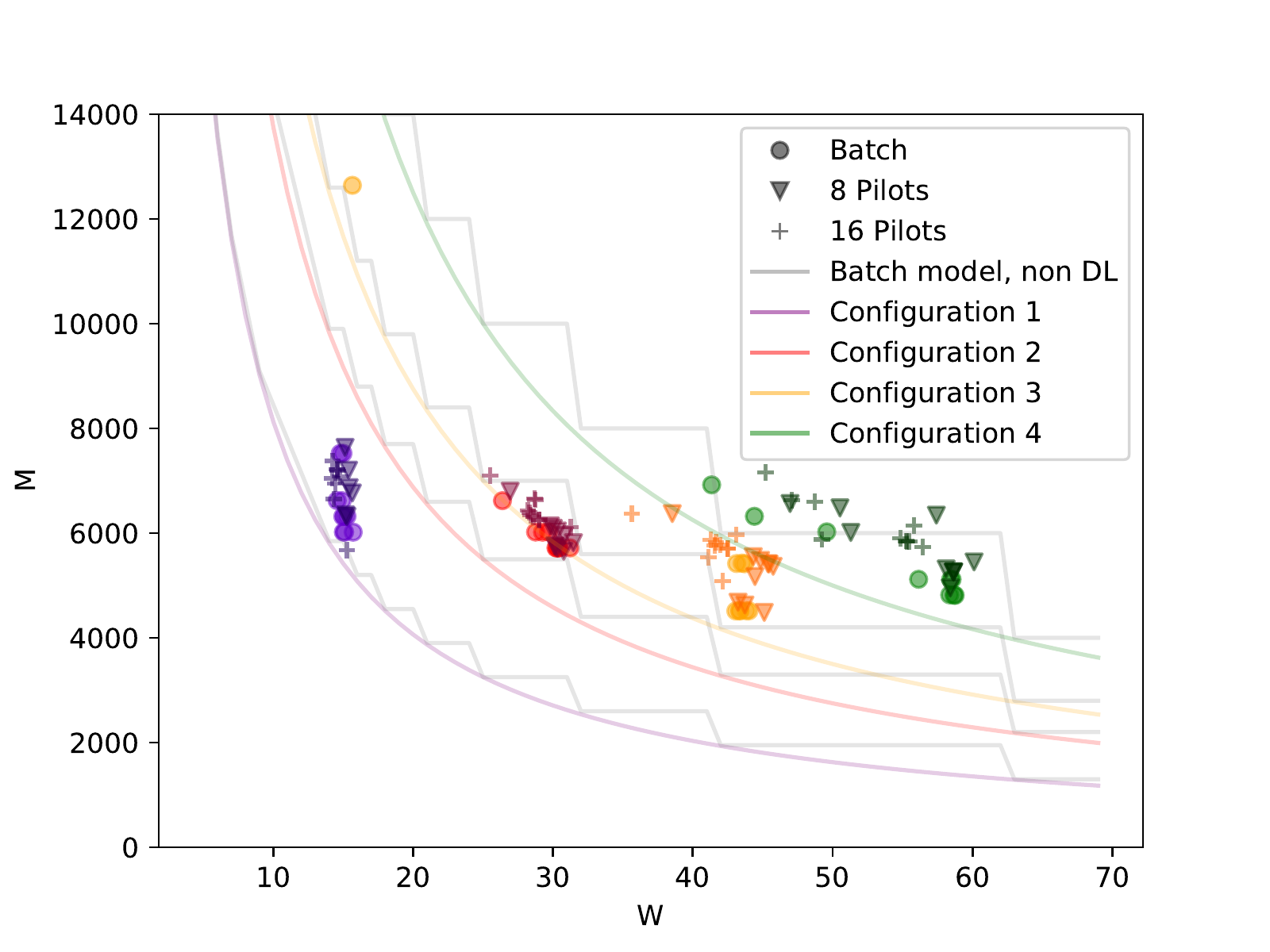}
            \caption[]%
            {{\small Beluga}}
            \label{fig:mwbeluga}
        \end{subfigure}
        \hfill
        \begin{subfigure}[b]{0.475\textwidth}
            \centering
            \includegraphics[width=\textwidth]{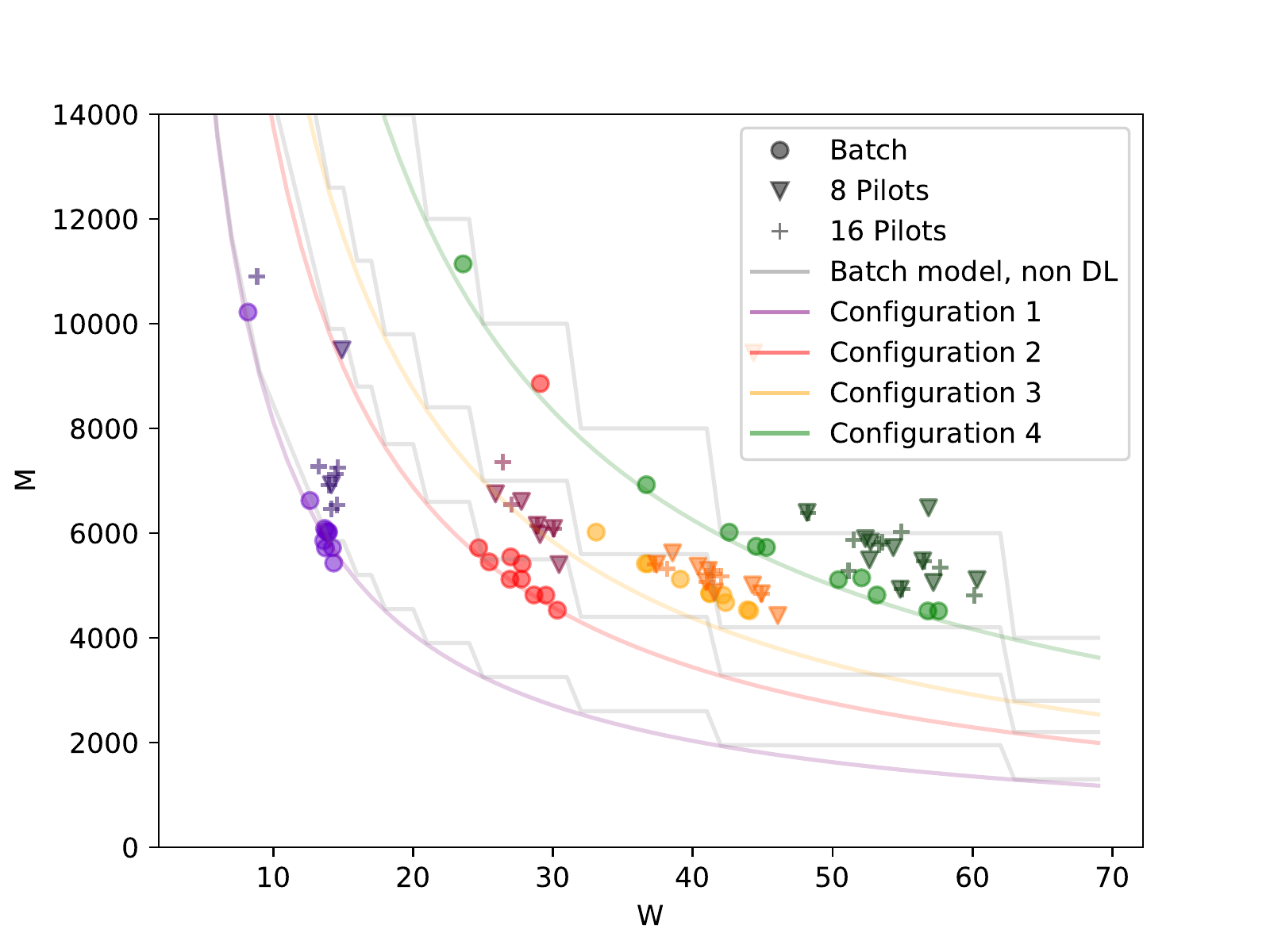}
            \caption[]%
            {{\small Cedar}}
            \label{fig:mwcedar}
        \end{subfigure}
        \caption[]
        {\small The relation between makespan and average number of workers as 
        calculated using Equation~\ref{eq:avgw}. Trendline denotes the expected makespan
        given the average number of workers given a certain configuration}
        \label{fig:mwall}
    \end{figure*}

In the case of pilots, multiple Slurm jobs, each with potentially different queuing times,
were used. Figure~\ref{fig:mwall} displays the makespan times for an average number of workers
given an experiment configuration, where the coloured line corresponds to the estimated makespan using 
our performance model. Makespan for the model was calculated as:
$$
M = \frac{125\times(d + 20)}{W}\times 10,
$$
where:
\begin{itemize}
    \item $125$ is the number of chunks BigBrain was split into
    \item $M$ is the makespan of the application, in seconds
    \item $W$ is the average number of Spark workers throughout the execution, computed as in Equation~\ref{eq:avgw}
    \item $d$ is the task delay, in seconds, associated with a given configuration
    \item $20$ is the average measured incrementation duration, in seconds, for a BigBrain chunk
    \item $10$ is the number of iterations
\end{itemize}
As can be seen in the Figure~\ref{fig:mwall}, both pilots and batch 
followed the general trend denoted by the model, which confirms that 
the system was behaving without major external sources of
overhead. However, pilots were consistently slower than batch for the same
number of average workers. This is particularly visible in Configuration 4
on B\'eluga (Fig. \ref{fig:mwbeluga}), and in Configuration 2, 3 and 4 on
Cedar (Fig. \ref{fig:mwcedar}). It means that some sort of overhead
impacted pilots but not batch. It can also be noticed that both batch and
pilots occasionally deviated from the model line, in particular
Configurations 2 and 3 in B\'eluga: this is due to the fact that our
application is not a divisible load, as the model assumes. Extending the
model beyond divisible loads is easy enough for batch and confirms that the
observed deviations from the model line come from this assumption (see gray
line in Fig.~\ref{fig:mwall}). The extended expression for pilots is more complex
though.

\begin{figure*}

        \centering
        \begin{subfigure}[b]{0.475\textwidth}
            \centering
            \includegraphics[width=\textwidth]{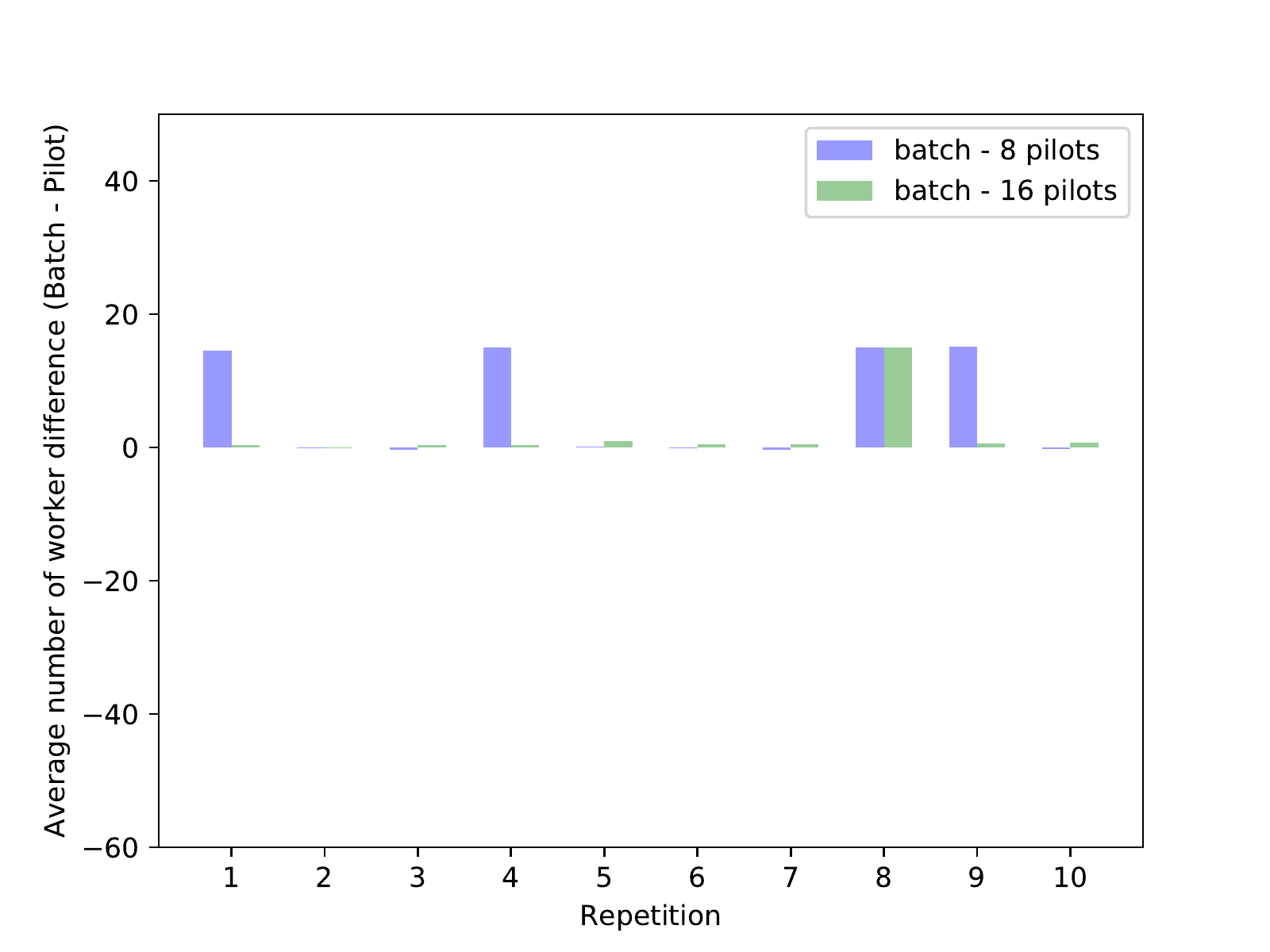}
            \caption[]%
            {{\small Configuration 1}}
            \label{fig:nwbeluga1}
        \end{subfigure}
        \hfill
        \begin{subfigure}[b]{0.475\textwidth}
            \centering
            \includegraphics[width=\textwidth]{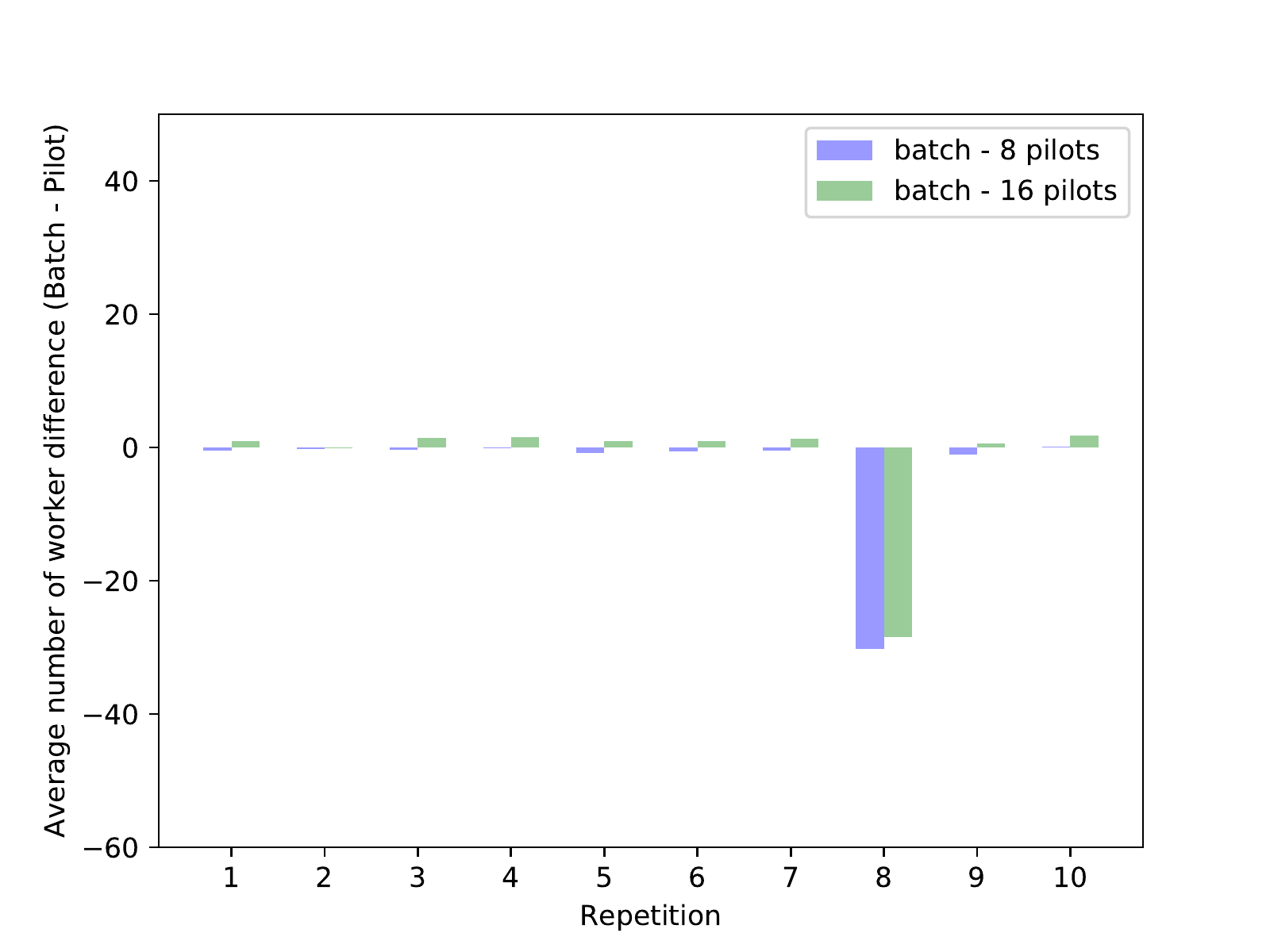}
            \caption[]%
            {{\small Configuration 2}}
            \label{fig:nwbeluga2}
        \end{subfigure}
        \vskip\baselineskip
        \begin{subfigure}[b]{0.475\textwidth}
            \centering
            \includegraphics[width=\textwidth]{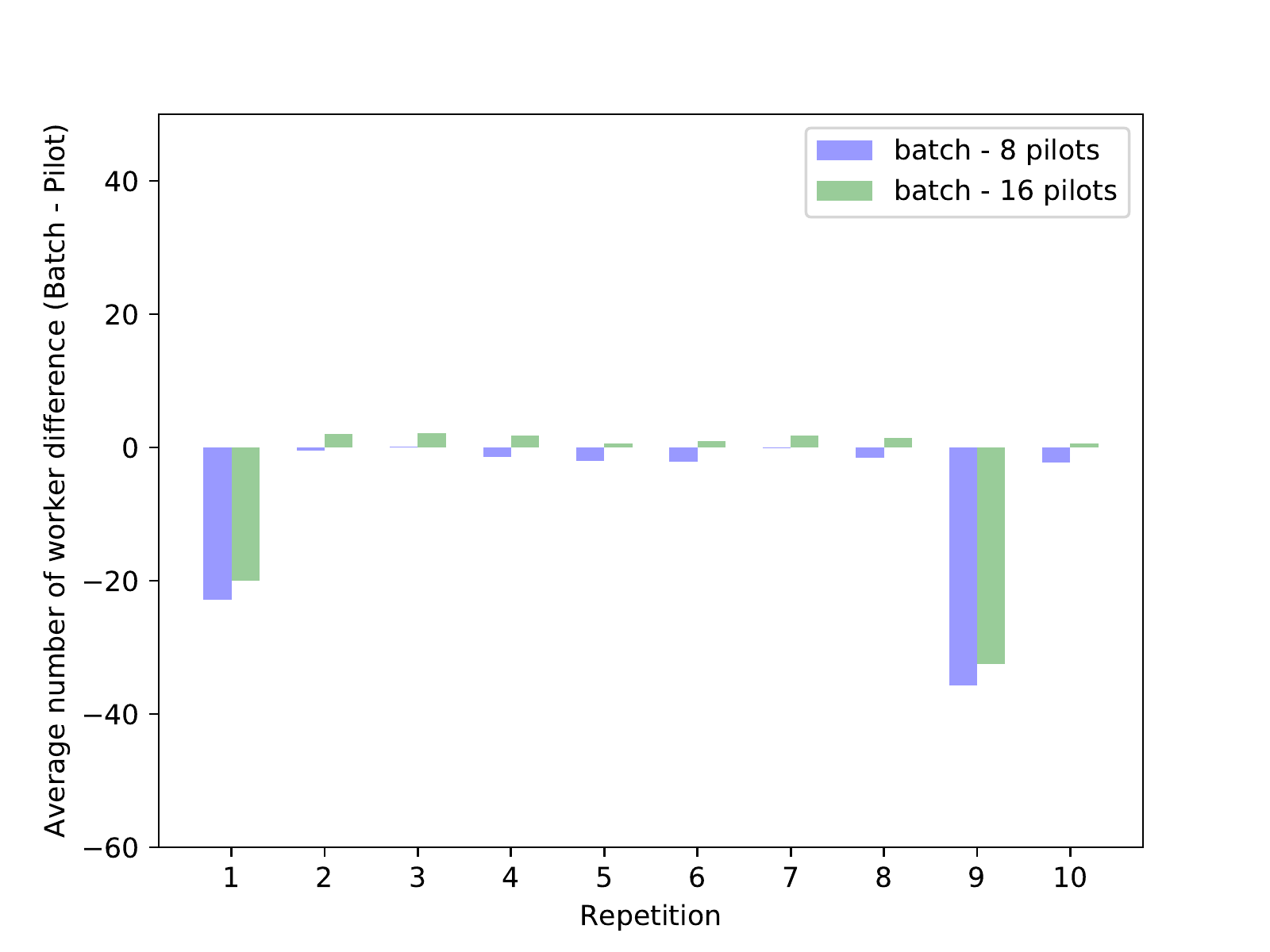}
            \caption[]%
            {{\small Configuration 3}}
            \label{fig:nwbeluga3}
        \end{subfigure}
        \quad
        \begin{subfigure}[b]{0.475\textwidth}
            \centering
            \includegraphics[width=\textwidth]{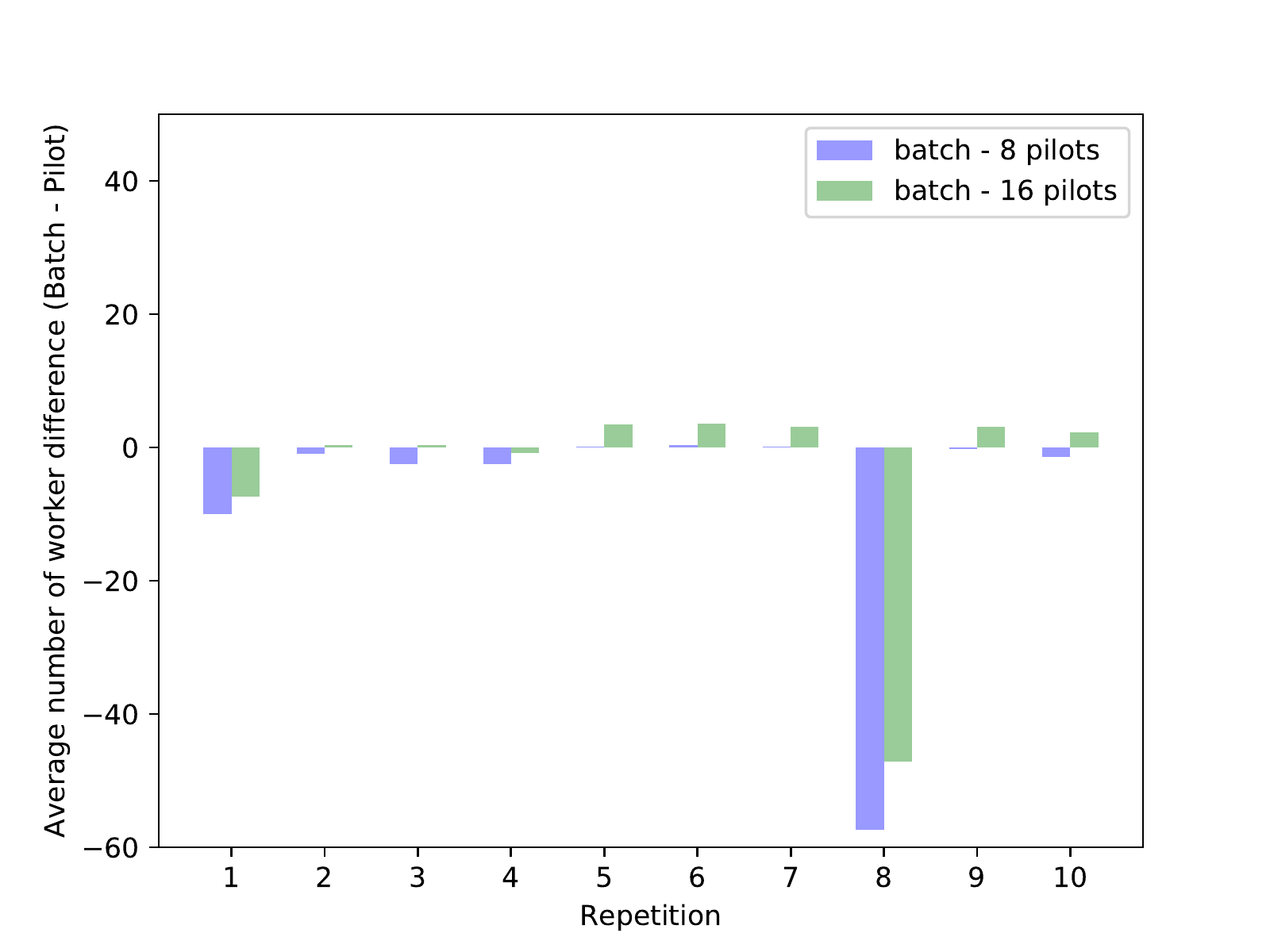}
            \caption[]%
            {{\small Configuration 4}}
            \label{fig:nwbeluga4}
        \end{subfigure}
        \caption[]
        {\small  The difference in average workers on B\'eluga between batch and pilots for all configurations and repetitions. Positive values mean
        that batch had more workers than pilots, that is, pilots did not improve queuing times.}
        \label{fig:nworkersbeluga}
    \end{figure*}

\begin{figure*}
        \centering
        \begin{subfigure}[b]{0.475\textwidth}
            \centering
            \includegraphics[width=\textwidth]{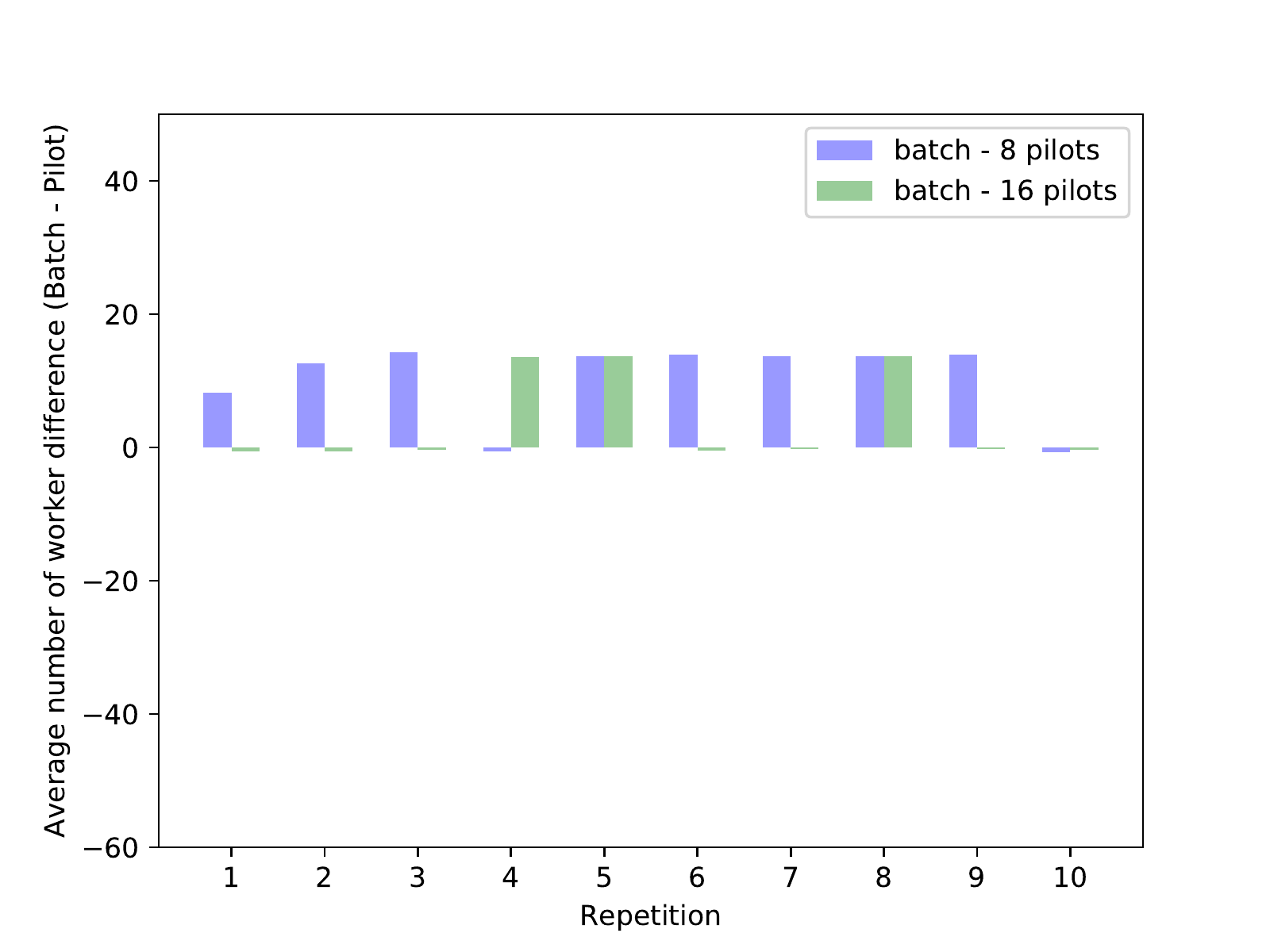}
            \caption[]%
            {{\small Configuration 1}}
            \label{fig:nwcedar1}
        \end{subfigure}
        \hfill
        \begin{subfigure}[b]{0.475\textwidth}
            \centering
            \includegraphics[width=\textwidth]{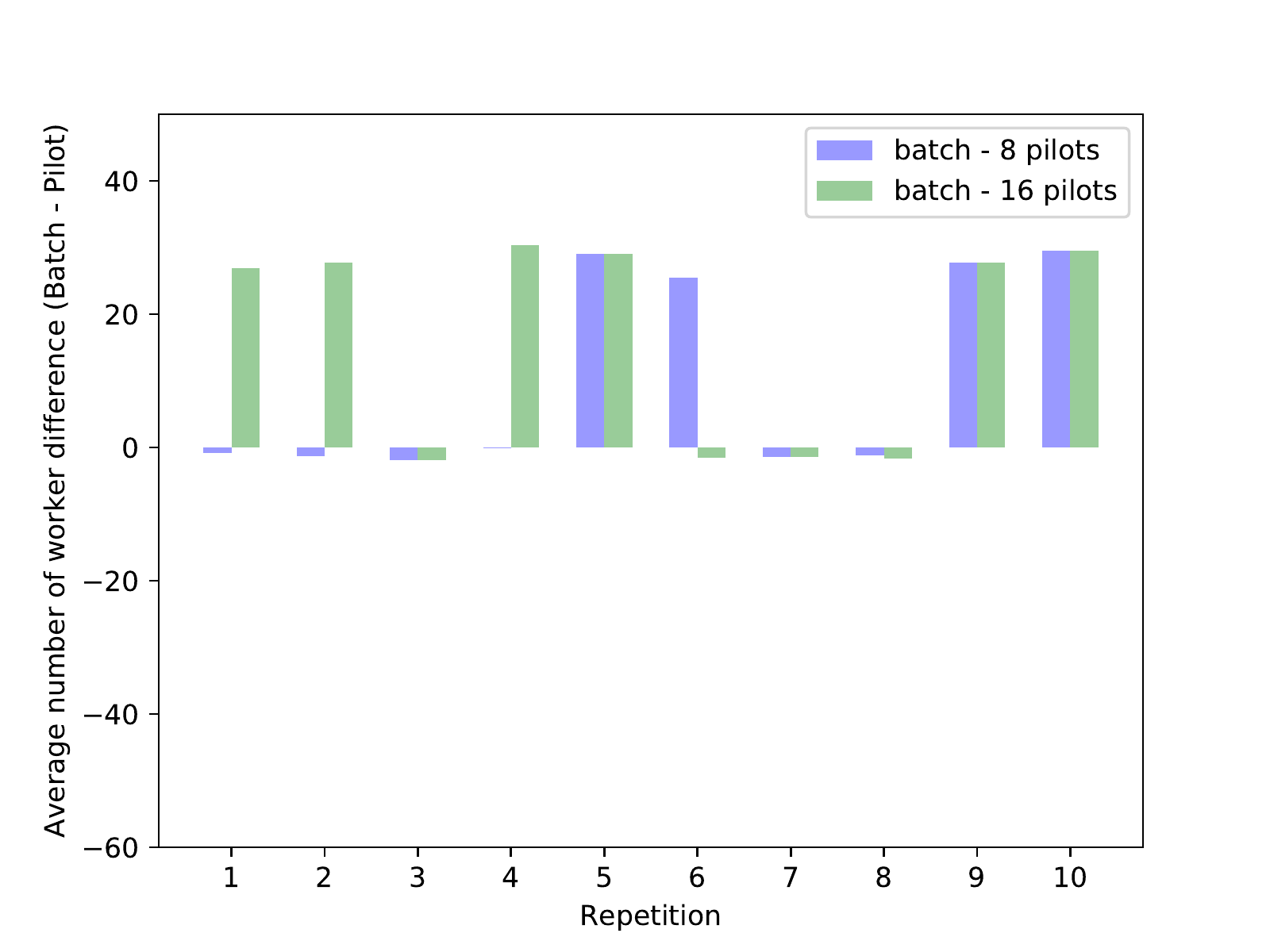}
            \caption[]%
            {{\small Configuration 2}}
            \label{fig:nwcedar2}
        \end{subfigure}
        \vskip\baselineskip
        \begin{subfigure}[b]{0.475\textwidth}
            \centering
            \includegraphics[width=\textwidth]{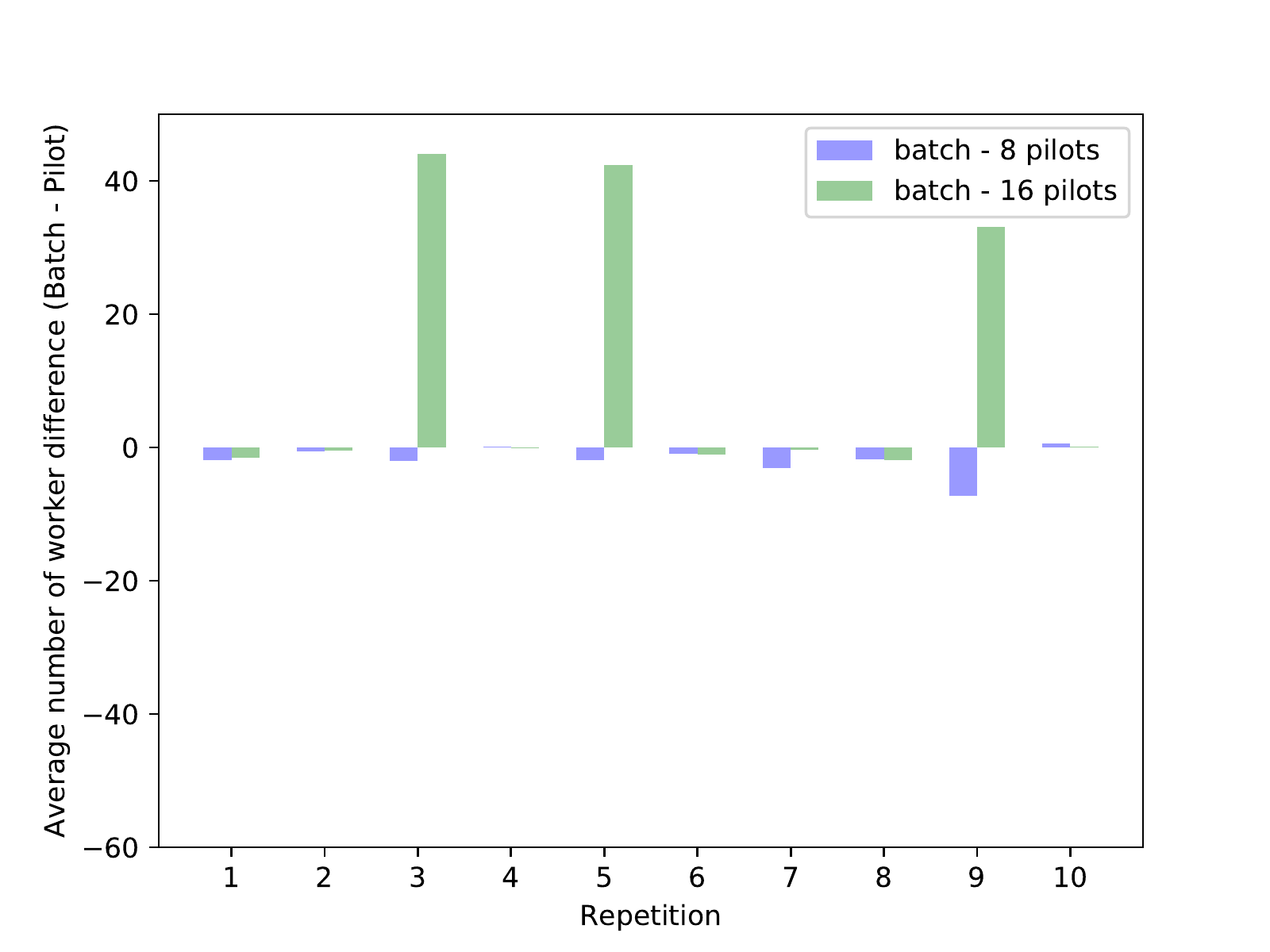}
            \caption[]%
            {{\small Configuration 3}}
            \label{fig:nwcedar3}
        \end{subfigure}
        \quad
        \begin{subfigure}[b]{0.475\textwidth}
            \centering
            \includegraphics[width=\textwidth]{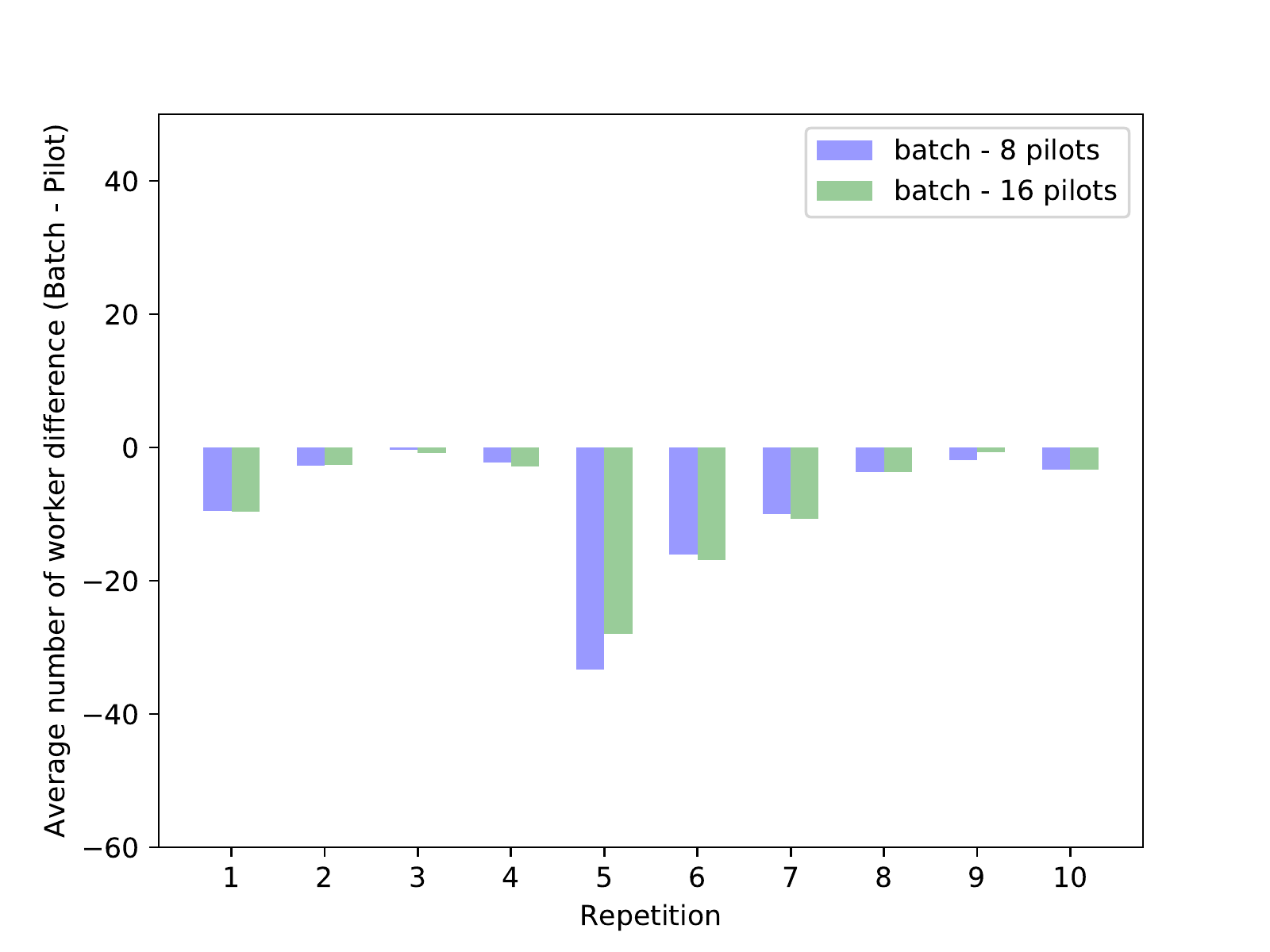}
            \caption[]%
            {{\small Configuration 4}}
            \label{fig:nwcedar4}
        \end{subfigure}
        \caption[]
        {\small The difference in average workers on Cedar between batch and pilots for all configurations and repetitions. Positive values mean
        that batch had more workers than pilots, that is, pilots did not improve queuing times.}
        \label{fig:nworkerscedar}
    \end{figure*}

When
comparing the average number of worker difference
(Figures~\ref{fig:nworkersbeluga} and~\ref{fig:nworkerscedar}), it can be
seen that (1) repetitions where pilots were faster than batch correspond
to repetitions where pilots had more workers, and vice versa -- this
confirms that performance differences are mainly coming from queuing time
differences, and (2) pilot average number of workers start to exceed batch
as the application requirements increase, that is, for Configuration 3 and
4.


\begin{table}                                                                    
    \centering                                                                       
    \begin{tabular}{c|c|c|c|c}                                                             
        {} & \multicolumn{2}{c}{B\'eluga} & \multicolumn{2}{c}{Cedar}\\
    \rowcolor{headcolor}                                                             
    Configuration & 8 pilots & 16 pilots & 8 pilots & 16 pilots\\                               
    \hline                                                                           
    1 & 0.949 & 0.932 & 0.724 & 0.874\\                                               
    2 & 0.988 & 0.916 & 0.836 & 0.826\\                                               
    3 & 1.458 & 1.369 & 0.941 & 0.964\\
    4 & 0.970 & 0.891 & 1.049 & 1.068\\
    \end{tabular}                                                                    
    \setlength{\belowcaptionskip}{-10pt}                                             
    \caption{Average speedup of pilots for each configuration}                                                    
    \label{table:speedup}                                                            
\end{table}

\section{Discussion}\label{sec:discussion}


Pilots do not appear to provide any performance advantage over batch
scheduling when deploying Apache Spark overlay clusters on HPC.
This can be due to a few reasons. The queuing time for batch and pilots were similar, 
which can be an indication that we were just lacking the right level of priority to 
have been scheduled immediately. In the instances where batch was found to be significantly slower,
it might have been case that we had obtained the necessary priority level but there
were not enough resources available to be scheduled.

It was also found that having 16 pilots was generally slower than 8 for the same
configuration. This was particularly the case on B\'eluga. We suspect this might be
related to the number of Slurm jobs permitted for a user on the backfill queue. For
all configurations, we ran batch, 8 pilots and 16 pilots concurrently. This would have
resulted in 25 Slurm jobs trying the access the queue concurrently. On
B\'eluga, the max number of jobs in the backfill queue is configured to a
total of 10 jobs. For 16 nodes, this would at best be only 62.5\% of the
pilots being scheduled at a given time if all resources are available.
The next 6 jobs would then have to wait 3 minutes before having an opportunity to enter the
backfill. Cedar, on the other hand, permits 40 user jobs to be backfilled at a given time.
Therefore, on Cedar, all concurrent Slurm jobs had an opportunity to be backfilled at
the same time, which may explain why this behaviour is more apparent on B\'eluga. Furthermore,
while both pilot configurations could be placed in B\'eluga's low memory nodes, batch requests
had to be place on B\'eluga's medium memory nodes. B\'eluga has 172 low memory nodes and
516 mid memory nodes. As pilots would get priority for the low memory nodes first and 
low memory nodes are less frequent that medium memory nodes, it is possible that this
may have increased the overall queuing times of the pilots. However, Cedar's basic
node has enough memory to for all batch and pilot configuration, therefore, this could not
have affected queuing time for Cedar.

Although queuing time is largely responsible for makespan variations, it is not
always entirely responsible for the difference. Sometimes there are errors related
to executors not properly starting. These occurrences lead to the 
application being processed entirely with a smaller number of available executors. As the task
delay added considers the maximum amount of parallel workers, functioning with less total workers
will significantly affect makespan. However, as seen in Figure~\ref{fig:mwall},
pilots are slower even with the same amount of average workers. A potential reasoning for this
may be the worker registration delay and the time required to transfer data
over after the application has already started. In batch scheduling, all workers are started in
parallel and the data is transferred automatically at the beginning of the pipeline as the maximum amount
of parallelism is available at driver start time. Pilots, on the other hand, may start workers in parallel,
but not necessarily all will be started at the same time (i.e. delays in starting the workers once resources
have been allocated). This means that the entire 
workflow may suffer the impacts of starting workers in sequence rather than in parallel. Moreover, there would also
be a delay with respect to transferring data from workers to newly added workers. Therefore,
it is only natural that pilots would have a bit more overhead than batch given the same
queuing time should pilots not all commence at once, despite transferring the same amount of data.
 Furthermore, the pilot scripts also
have some startup overhead as all pilots start a master, a worker (registered to the main
master) and attempt to start a driver. Batch needs only start a single master and attempts to start
a driver only once.

The motivation to use pilots, however, is dynamic scheduling on HPC clusters. While we
investigated the queuing time differences with respect to differences in available resources,
the walltimes were kept static. Had walltime been underestimated to the point where the Slurm jobs
would terminate prior to application completion, pilots might have been preferable to batch
as batch scheduling has no mechanism to restart. Even if such a mechanism was in place, the entire
cluster would be shutdown and need to restart, which has additional overheads. With pilots, as long as 
there was at least one pilot alive, it would be possible to maintain the cluster and add additional pilots.
Spark provides built-it fault-tolerance for not only the workers, but the master and driver as well. However,
it is also likely that with faster queuing times, all resources would be allocated at the same time. If walltime
were to be underestimated, in this case, checkpointing and restarting the entire application from the last
checkpoint would be the only option. This too has limitations as the last checkpoint may not be recent,
necessitating significant amounts of recomputation. It
might even be desirable to have many small resource requests in such a scenario, as it becomes less likely for all pilots to start at once as they become more numerous, as can be seen with 16 pilots when compared to 8. This would allow
for more chances for the application to reach completion without failing entirely.

Many failures can be found, particularly with regards to the pilots. Most
of these have been a result of Spark failing to start properly. The driver
either would not return a submission ID after being started, or would
remain in \texttt{SUBMITTED} status until the walltime expiration, having
not processed any data. As these errors become fewer with increased
resources, it is possible they are entirely related to lack of resources.
Furthermore, it may also explain why batch submission is more frequently successful. Not
only are all the workers that were requested available when the pilot
begins running, but batch also has one extra executor available thanks to the driver
process running directly on the process which had created it, rather than on a
worker process. Conversely, it is not only the pilot applications that
fail, the batch applications do experience some failures as well. It can be
seen in Figure~\ref{fig:makespansbeluga} that the batch application failed
twice: once in Configuration~2 and the other in Configuration~4. In both
cases, the reason for failure was the same: lack of available resources
despite all executors having been assigned. These failures might be due to
running on a shared HPC cluster. Either the resources were down, or other users might
have been using them, resulting in the resources being unavailable to our
application.

Running a Spark cluster atop of a Slurm allocation complexifies debugging, particularly if the application runs in cluster mode.
Spark worker logs, unless saved to network storage, become unavailable after program execution.
Although the worker UI is available during program execution, a user must manually create an SSH tunnel
to each worker node to gain access to them.
An option is to save the to a shared network storage for future access. This effectively slows down
the application as the network file system is typically a slower storage than local disk and RAM.

When running in client mode, the driver output ends up it standard output/error. 
However, in cluster mode, the driver output is found stored within
the worker logs, which means the driver logs are also inaccessible without 
worker logs being written to network storage. Some HPC clusters
may have a file limit for users, meaning that the user must be sure that these 
logs coupled without other stored data does not exceed the file
limit.  

\section{Conclusion}\label{sec:conclusion}
Pilots do not provide significant advantage over batch systems to run
Apache Spark applications on HPC clusters. This is affected by the overhead of dynamically
adjusting the cluster size and the fact that queuing time of the batch
applications took as long as the pilots. These results, however, are to an
extent cluster-dependent. It is expected that as the number of resources
requested increases, the speedup provided by pilots will increase.
Furthermore, users executing on a smaller cluster may experience greater
speedups using the same resource configuration. Our resource requirements
and cluster size, however, match typical configurations currently available
to scientific communities.

Although obtained with Apache Spark, this result is likely to generalize to
other implementations of overlay clusters. Conversely, our experiments do
\emph{not} imply that pilot jobs are not useful on HPC clusters in the
absence of overlay clusters. Most pilot job implementations in fact include
an overlay scheduler, which allows for various types of optimization
including dealing with fine task granularity or enforcing data locality.
Pilot jobs naturally remain useful in that respect.

Pilots are meant to allow for underestimation of resources, particularly
walltime, which would not be known to the user prior to execution. In its
current state, \texttt{SPA} is not able to evaluate that functionality.
Further experiments on master fault-tolerance, driver recovery, and
checkpointing will need to be conducted to determine if pilot jobs are even
a viable solution for Spark applications running on HPC.

Currently, as client mode is not available in Spark Standalone clusters for
PySpark applications, using pilots on HPC would be limited to non-Python
applications until there is a solution for Standalone mode.

\section*{Acknowledgement}

We warmly thank Compute Canada and its regional centers WestGrid and Calcul
Qu\'ebec for providing the infrastructure used in our experiments.

\bibliographystyle{IEEEtran} 
\bibliography{biblio}
\end{document}